\documentclass[twocolumn,aps,pre,superscriptaddress,floatfix]{revtex4-1}
\usepackage{graphicx}
\usepackage{epsfig}
\usepackage{thumbpdf}
\usepackage{amsfonts}
\usepackage{times}     
\usepackage{float}
\usepackage{indentfirst}   %
\usepackage{amsmath}
\usepackage{epstopdf}
\usepackage{textcomp}
\usepackage{tocvsec2}
\usepackage{epsfig}
\usepackage{xcolor}
\usepackage{comment}
\usepackage{CJK} 

\usepackage[section]{placeins}
\makeatletter\AtBeginDocument{%
     \expandafter\renewcommand\expandafter\subsection\expandafter
       {\expandafter\@fb@subsecFB\subsection}%
     \newcommand\@fb@subsecFB{\FloatBarrier
     \gdef\@fb@afterHHook{\@fb@topbarrier \gdef\@fb@afterHHook{}}}
     \g@addto@macro\@afterheading{\@fb@afterHHook}
     \gdef\@fb@afterHHook{}
  }
\usepackage[colorlinks=true,linkcolor=black]{hyperref} 

\newcommand{\be}{\begin{equation}}
\newcommand{\ee}{\end{equation}}

\begin{document}
\begin{CJK}{UTF8}{gbsn}

\title{Exploring the formation dynamics of affective polarization by considering a coupled feedback}

\author{Peng-Bi Cui (崔鹏碧)}\email{cuisir610@gmail.com}
\affiliation{International Academic Center of Complex Systems, Beijing Normal University, Zhuhai, 519087, China}

\begin{abstract}
Polarization issue is generally subject to ideological polarization and affective polarization. In particular, affective polarization usually  accelerates the polarization process and transform social interactions into a zero-sum game. Yet, a wide array of existing
literature have not provided valid ways to make distinction between them. Therefore, the mechanism contributing to the rise of affective polarization still remain unclear, as well as its unique emergent dynamics. To address this issue, this study introduces the coupled feedback between opinions and response susceptibility to a attraction-repulsion model which takes account into three parameters: interaction strength, response susceptibility and tolerance to others. The model features phase diagrams of global consensus, affective polarization, and ``harmony with diversity" states. The simulations on time-varying and static social networks show that intermediate parameter ranges yield a global convergence, as one integrated cluster collapsing and converging towards a uncertain moderate position after long-time persistence. Overall, the simulations reveal that the feedback essentially offers a counterforce to establish an inversion between global convergence and ``harmony with diversity". Remarkably, strengthening feedback may facilitate polarization by driving the system priorly self-organize into one integrated cluster which then gradually approaching polarization, especially for low tolerance and strong interactions, and the step-like dynamic behaviors of opinion entropy suggest the occurrence of dynamic equilibrium. The above phenomena have never been unearthed before, which can be regarded as unique dynamics features of affective polarization. For the first time, this study attempts to offer a useful approach to the micro foundations of affective polarization, and the results guide us how to avoid the dilemmas from this polarization.
\end{abstract}

\maketitle

\section{Introduction}
\label{sec:introduction}
Polarization is here conceived as a process of growing constraint in people's viewpoints, and further emergence of alignments along multiple or even opposite lines of potential disagreement~\cite{Axelrod1997,Bramson2017}. Nowadays the world is being perplexed by the growth of polarization~\cite{Guilbeault2018,Allcott2020}. It is thus increasingly a concern to understand the nature and causes of polarization and how to prevent the public from reaching dangerous degrees of polarization~\cite{Castellano2009,Sobkowicz2012,Moe2012}.

There exist at least two kinds of polarization to be easily carried to extremes which can undermine the foundation of social order: ideological polarization (IP)~\cite{Lelkes2016} indicating an opposite political view among elites, and affective polarization (AP) whereby individuals develop negative attitudes toward members of the opposing party and become entrenched~\cite{Iyengar2012,Lauka2018,Iyengar2019,Boxell2020,Reiljan2020,Wagner2021}. IP and AP have been rising during the last decades~\cite{Abramowitz2008,Boxell2020}. Furthermore, IP refers to that political views are widely dispersed and heterogeneous or bimodal among elites. Unlike IP largely driven by rational self-interest motivations~\cite{Lelkes2016,Axelrod2021,Grossman2021,Perrings2021,Vasconcelos2021}, AP itself generally suggests a strong affective decision-making with a coupling relationship between individuals opinions and their subsequent responses~\cite{Iyengar2019}. This coupling relationship essentially creates a negative feedback mechanism, which is widely existed in empirical cases~\cite{Iyengar2019}. For example, individuals with higher levels of out-party animosity report taking a more strong stand against those from the other party (for example, they would not even receive any suggestion from those from the other party)~\cite{Lelkes2017}, and thus more difficulty to reposition, which is regardless of economic self-interest. The adverse effects of AP can spill over from the political area to other issues such as COVID-19 vaccines~\cite{Woolhandler2021}, gun-control~\cite{Sunstein1999} and abortion~\cite{Dimaggio1996}, which is much more harmful than IP being. Therefore, AP might have unique dynamics, behind which the mentioned micro negative feedback can be naturally a potential responsible mechanism.

Up to now, a large bodies of existing studies have explored various mechanisms which are responsible for clustering of beliefs and polarization especially IP in complex social systems~\cite{Castellano2009,Kossinets2009,Bramson2017,Vasconcelos2019,Baumann2020,Santos2021,Axelrod2021,Chu2021,Jusup2022}, and the mutually reinforcing relation between IP and AP has also be identified. For example, Americans increasingly dislike those of the other party because of intense struggles between Democrat and Republican~\cite{Mason2016,Luttig2017,Iyengar2019}, which conversely gives rise to an increasing IP among the US public, rising social inequality and conflicts~\cite{Dellaposta2020}. Furthermore, polarization among elites has become mainly affective rather than ideological. The above realistic problem suggests a call for further explorations to identify the unique dynamic features of AP, so as to develop more effective preventing measures to slow or even stop the accelerating polarization process, which has nevertheless received too little attention in the growing literature. Due to the positive feedback, however, the growth of AP seems to have multiple contributing factors, including economic adversity, racial animus, culture values, religions, gender polarization and a range of other socioeconomic factors~\cite{Schaffner2016,Sides2016,Luttig2017,Mitrea2021}, making the exploration of unique features of AP being arduous. Since the general micro foundation of AP dynamics remains unclear. It leads to the following questions need to be addressed: What the micro foundation of AP dynamics, and the unique dynamic features of AP are? How the micro mechanism governs the AP dynamics, together with other potential factors such as interaction strength, response susceptibility, and tolerance level?

In order to address these questions arising from the lack of micro foundation of AP, we adopt a novel attraction-repulsion
model (ARM) that captures the coupled feedback between individuals' opinions and their response, which essentially belongs to negative feedback class. The proposed model constructs an adaptive-system perspective on the effects of the rules on emergent dynamics of not only AP and global census (GC), but also ``harmony with diversity" (HD) state which has been recently identified~\cite{Cui2023}. Note that the opinion updating rules in our model paradigm are based on a simple assumption that individual's attraction to or repulsion from others is only governed by the opinion dissimilarity between them, rather than the positions themselves. In addition, interaction strength, response susceptibility and individual tolerance, which have been viewed as vital regulation factors to be considered. We are concerned how the negative feedback mechanism govern the emergent dynamics of AP, together with the three vital factors. 

The discoveries of this study include: 1) The identification of conditions under which a population approaches convergence into a moderate position. 2) The identification of the conditions under which the population becomes highly polarized with asymmetric opposite camps or even extremely polarized with symmetric opposite extreme camps, and under which it enters a desired HD state with a stable integrated opinion cluster centering on neutral point. 3) The transitions between the three states, as well as the triple point. 4) Opinions of the majority may firstly self-organizes into one integrated cluster which then approaches the boundary of ideological space because of counterforce exerted by the negative coupled feedback arising from frequent opinion exchange within the cluster. 5) The system collapses into a global convergence towards a moderate position after a long-time persistence of one integrated opinion cluster. 6) The remark that strengthening feedback can undoubtedly increase the likelihood of AP.
For the first time, our study proposes a basic model framework to explore the unique dynamic features of AP.

The paper is organized as follows. In Sec.~\ref{sec:model}, we formally define our opinion evolution model, along with the method to identify the regions of different states. In Sec.~\ref{sec:results}, we simulate our model, and give results, as well as descriptions, attempts to identify the unique dynamics features of AP. 
We conclude with a summary of the results and an outlook for future studies in Sec.~\ref{sec:conclusion}.

\section{Model}
\label{sec:model}
The model considers a population of size $N$ where each individual is typified by an opinion $x_{i}(t)$ at time $t$, which is a real number in the interval $x_{i}(t)\in[-10,~+10]$. We adopt the opinion model proposed by Ref~\cite{Cui2023}, where the updating rules of individual's opinion $x_{i}(t)$ is formulated by the following equations:
\begin{eqnarray}
\hspace{-.5cm} \dot{x}_{i}(t) & = &
\begin{cases}
A\tanh(\alpha_{i}D_{ji}(t))\quad \text{if}~|D_{ji}(t)|<T_{i}; \\ 
A\tanh(\alpha_{i}\sigma(D_{ji}(t))(T_{i}-|D_{ji}(t)|))\quad \text{if}~|D_{ji}(t)|\geq T_{i}. 
\end{cases}
\label{eq:asfunztion}
\end{eqnarray}
$D_{ji}(t)=x_{j}(t)-x_{i}(t)$ denotes the opinion distance between $i$ and $j$ at time $t$. $T_{i}$ is the tolerance threshold of individual $i$. $\alpha_{i}$ can be interpreted as response susceptibility of the individual. In more detail, $\alpha_{i}$ positively associates with the extent to which individual $i$ is passionate or sensitive, i.e., response susceptible to be socially influenced. It is obvious that nonlinear shape of the influence function $tanh(x)$ is controlled by $\alpha$. While $A$ quantifies interaction strength, which is actually the upper bound of opinion shift driven by each interaction, implying that the influence exerted by individuals on others is capped. $\sigma(D_{ji}(t))$ extracts the sign of $D_{ji}(t)$. 

We firstly consider a population where the connections of each individual are dynamic and formulated by activity driven (AD) model~\cite{Liu2014,Perra2012,Moinet2015,Baumann2021}. In more detail, $k_{i}(t)$ denotes the number of interactions individual $i$ has at time $t$. It thus generates a temporal network formulated by the temporal adjacency matrix $A_{ij}(t)$, where $A_{ij}(t)=1$ if individual $i$ owns one connection with individual $j$, otherwise $A_{ij}(t)=0$. Within the AD model framework, each individual $i$ randomly connects $k_{i}$ individuals from the population at each time step. $k_{i}=\sum^{N}_{j}A_{ij}(t)$ is thus satisfied throughout the simulation. The opinion evolution is actually coupled to an underlying temporal network. In reality, the empirical statistics that interaction activities of people are generally heterogeneous~\cite{Perra2012,Moinet2015,Baumann2020}. We thus further assume the interactions are extracted from a power-law distribution $p(k)\sim k^{-\gamma}$. 

The proposed model incorporates the micro negative feedback mechanism by which individuals owning more extreme opinions becomes more stubborn and less sensitive, and thus more difficulty to reposition. The micro feedback mechanism with regards to affective decision-making is based on individual level, and related to intrinsic preferences. On the basis of the basic model defined by Eq.~\ref{eq:asfunztion}, we additionally assume the following feedback function regarding either interaction susceptibility $\alpha$, using individual opinion as the variable:
\begin{eqnarray}
\label{eq:alphafunction}
\alpha_{i}(t)=\eta x^{-\beta}_{i}(t),
\end{eqnarray}
where $\beta$ quantifies the strength of the negative feedback. The larger $\beta$, the stronger the feedback becomes, and as a consequence, even moderate individuals will not change their position easily. The feedback function captures the coupled relationship between individuals' opinion and their preference. This means that the present model includes intrinsic preferences for a specific opinion. Varying $\beta$ may generate different evolution dynamics, facilitating an understanding about the effects of the micro feedback on the emergent dynamics of global polarization, global convergence and ``harmony with diversity" (HD) states.

Next, we extend the proposed model from AD network to the static network where the interactions among individuals are fixed, such as the parts of the online social network Facebook. In such case, $k_{i}$ corresponds to the number of edges that individual $i$ stretches to its neighbors, and the individual is represented by a node of the networks. Therefore, the ones with which each individual interacts keep unchanged. 

In numerical simulations, the size of the employed time-varying network is $N=1000$, where $\gamma=2.1$. The control parameters of the present model are $A$, $\beta$, $T_{i}$, where we assume that individuals have uniform attributes $\alpha_{i}=\alpha$, $T_{i}=T$ and $A_{i}=A$. For simplicity, we set $\eta=1.0$. The final results are obtained from $N_{r}=100-500$ independent realizations, after at least $500$ time steps. Before starting each realization, the initial opinion of each individual is independently and randomly sampled from the interval $[-1.0,~1.0]$. Then at each time step $t$, opinion evolves as following: (i) In random order, each individual $i$ randomly choose $k_{i}$ new neighbors out of all individuals. While the neighbors of $i$ keep unchanged in the static social media network. (ii) Then, $i$ compares its opinion with each neighbor, attempting to update its opinion with adaptive response susceptibility $\alpha(x_{i}(t-1))$ defined by Eq.~\ref{eq:alphafunction}.

We exhibit the polarization dynamics through the polarization degree of a population, which is measured by the standard deviation (SD) in opinions $SD(x_{0} ,..., x_{N})$. We measure the opinion diversity by calculating the opinion entropy ($S$) of the population: $S =\sum^{x_{max}}_{x_{min}}x\rho_{x}$, so as to capture the emergence of HD state. $\rho_{x}=\frac{N_{x}}{N}$ is the density of individual owning opinion $x$, and $N_{x}$ denotes the population of opinion $x$. Therefore, $x_{min}=-10$ and $x_{max}=10$ in our model. 
Larger $SD$ indicates increased degree of polarization. For example, the minimum of the polarization is zero i.e. $SD_{min}=0$, corresponding to GC state. Otherwise, the system will get extremely polarized if $SD$ is large or equal to $SD_{max}=10$. 

Inspired by the phase-identification method developed in Ref~\cite{Cui2023}, we still numerically estimate the boundaries between different sates by means of the susceptibility of not only $S$ but also $SD$. 
\begin{eqnarray}
\chi(S) & = & \frac{\sqrt{\langle S^{2}\rangle-\langle S\rangle^{2}}}{\langle S\rangle}, \label{eq:varianceS}
\end{eqnarray}
\begin{eqnarray}
\chi(SD) & = & \frac{\sqrt{\langle SD^{2}\rangle-\langle SD\rangle^{2}}}{\langle SD\rangle}, \label{eq:varianceSD}
\end{eqnarray}
where $\langle S\rangle$ ($\langle SD\rangle$) is the ensemble average of $S$ ($SD$), which can be obtained by averaging $S$ ($SD$) from $N_{r}$ independent realizations. $\langle S^{2}\rangle$ ($\langle SD^{2}\rangle$) is the secondary moment of the ensemble distribution. One can further identify the boundaries between GC, HD and AP states according to the principle that $\chi(S)$ and $\chi(SD)$ exhibits a peak value at the boundary.

\section{Results}
\label{sec:results}

\begin{figure*}
\centering
\includegraphics[width=\linewidth]{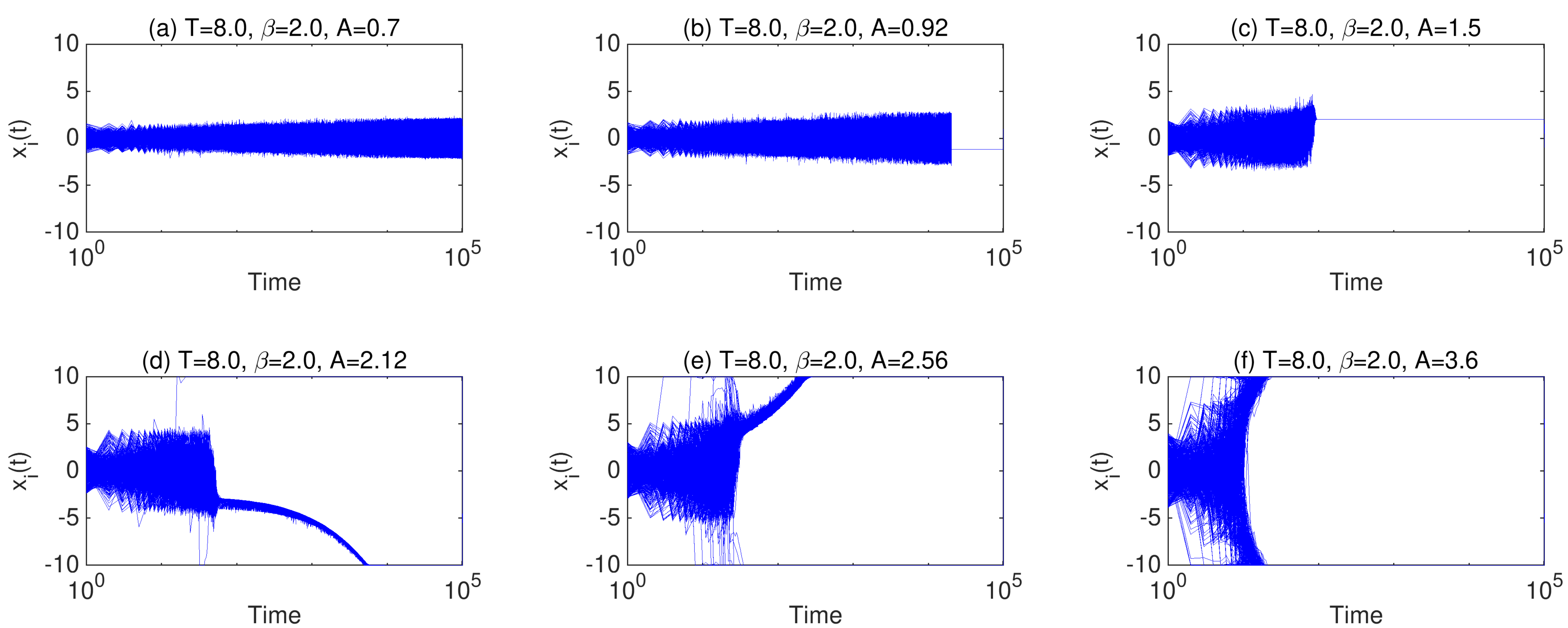}
\caption{
Temporal evolution of the individuals' opinions. The six representative cases are also indicated by the dark green dots in Figs.~\ref{fig:figure4}(a1) and (a2) for a more explicit presentation. The values of the parameters are listed in the titles of the subplots. 
}
\label{fig:figure1}
\end{figure*}

\begin{figure*}
\centering
\includegraphics[width=\linewidth]{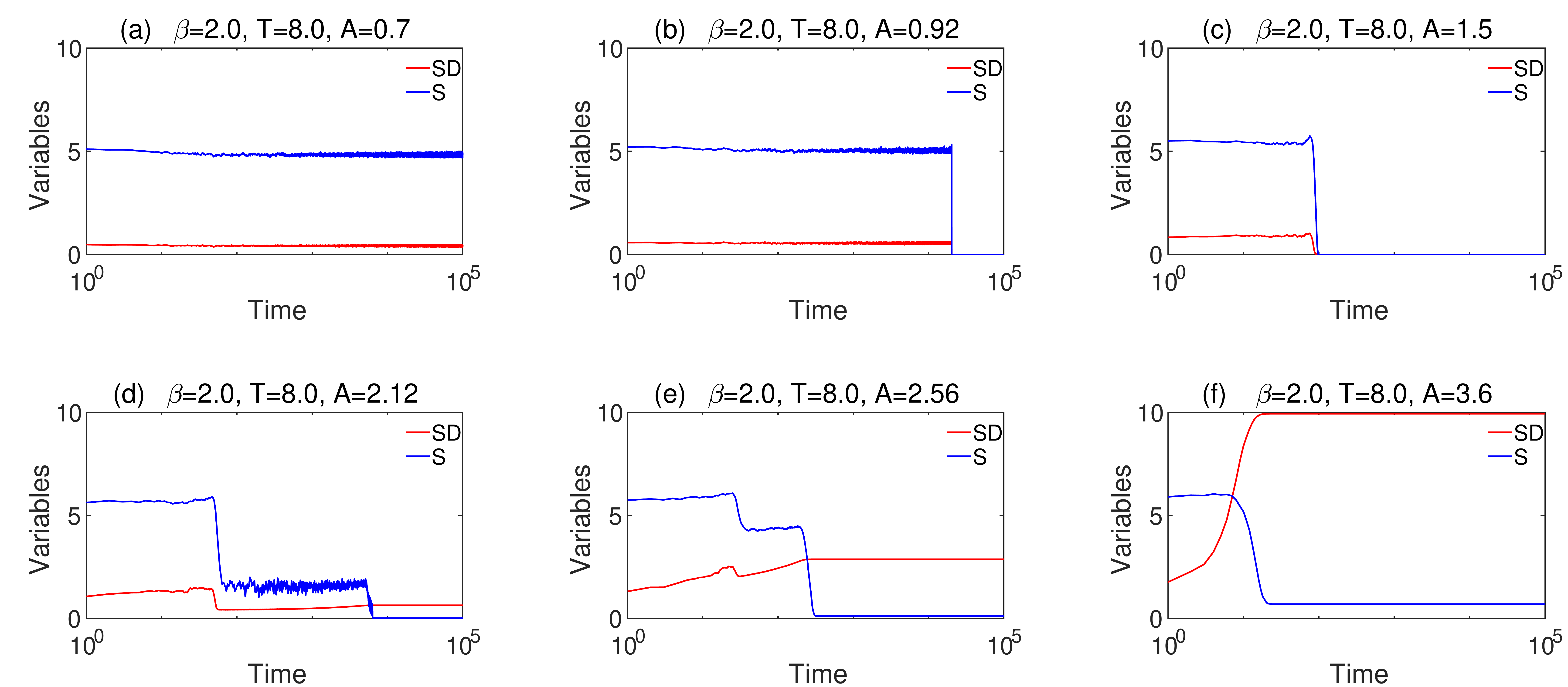}
\caption{
Temporal evolution of $SD$ and $S$. The representative six cases are also indicated by the dark green dots in Figs.~\ref{fig:figure4}(a1) and (a2) for a more explicit presentation. The values of the parameters are the same as considered in Fig.~\ref{fig:figure1}. 
}
\label{fig:figure2}
\end{figure*}

\begin{figure*}
\centering
\includegraphics[width=\linewidth]{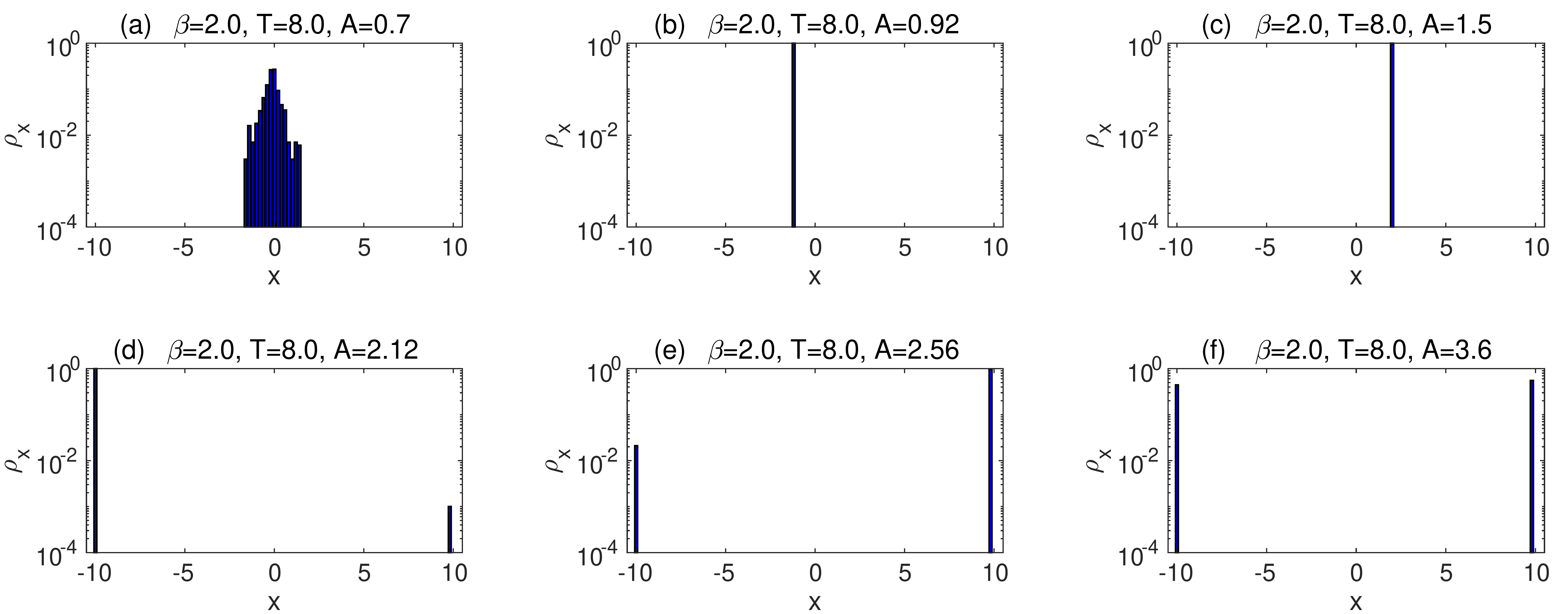}
\caption{
The final opinion distributions of the population, where the values of parameters are listed in the titles of the subfigures, corresponding to the six cases illustrated in Figs.~\ref{fig:figure1} and \ref{fig:figure2}. 
}
\label{fig:figure3}
\end{figure*}

\begin{figure*}
\centering
\includegraphics[width=\linewidth]{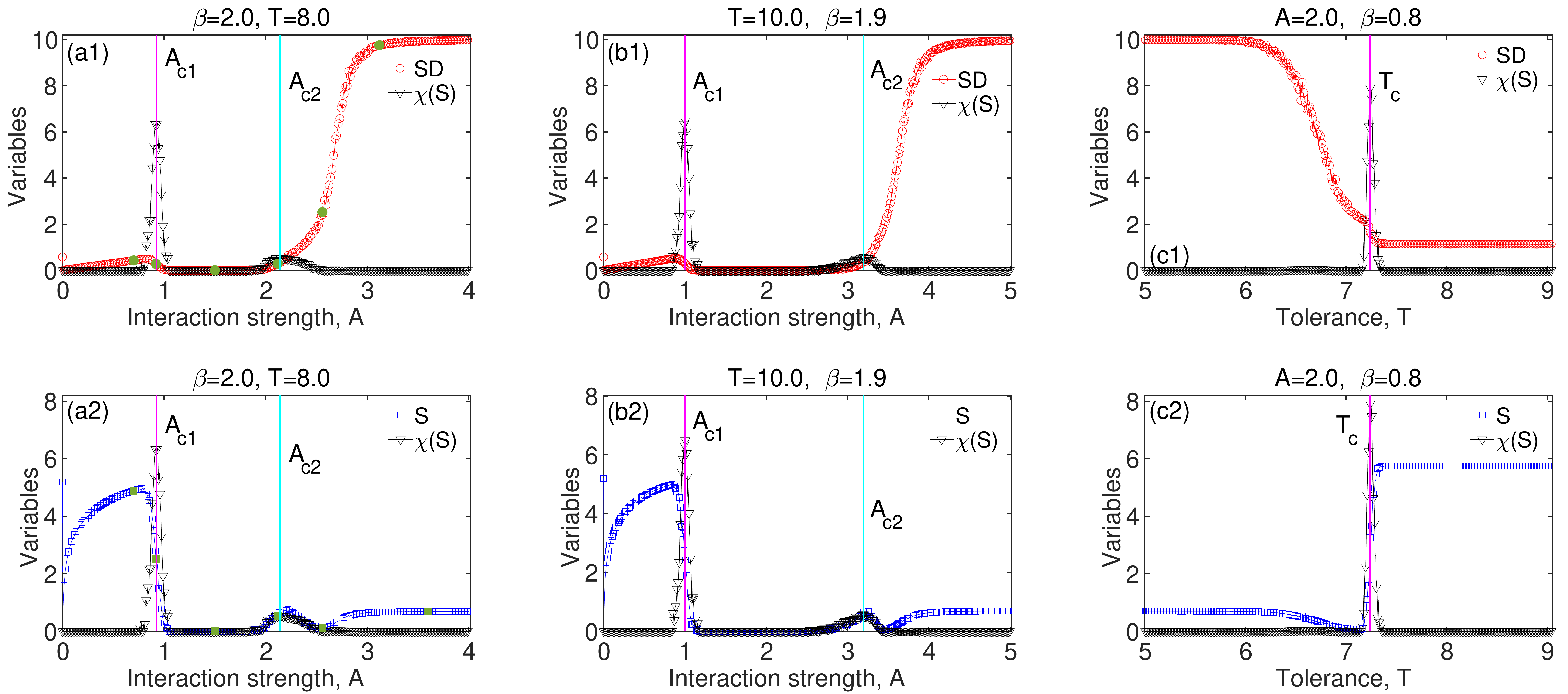}
\caption{The dependence of $SD$ (red circles), $S$ (blue squares), $\chi(S)$ (black triangles) on $A$ or $T$ in time-varying networks.  In (a1) and (a2), the six dark green dots correspond to the subplots of six parameter combinations in Figs.~\ref{fig:figure1} to ~\ref{fig:figure3}. In (a) and (b), pink vertical lines label the positions of $A_{c1}$ at which GC state begins to appear, while light blue vertical lines indicate the positions of $A_{c2}$ at which GC state starts to vanish, and shift to AP state. While we can observe the threshold $T_{c}$ that denotes a shift from AP to HD. The values of parameters are correspondingly listed in titles of subfigures. 
}
\label{fig:figure4}
\end{figure*}

Figs.~\ref{fig:figure1} and \ref{fig:figure2} present the temporal evolution of the individuals' opinions, polarization degree $SD$ and opinion entropy $S$ for six different parameter conditions, respectively. Corresponding final opinion distributions are illustrated in Fig.~\ref{fig:figure3}. We can observe that as interaction strength $A$ increases, the system firstly shifts from HD to GC, then from GC to AP state. More in detail, the presence of opinion-based feedback can facilitate a stable integrated opinion cluster centering on neutral point (see Fig.~\ref{fig:figure1}(a)), along with stable considerable $S$ and low-level $SD$ (see Fig.~\ref{fig:figure2}(a)). Since individuals are allowed to remain greatly sensitive to diverse views of others in spite of opinion distance being slight (see Fig.~\ref{Sfig:figure1}(a)). This occurs frequent opinion exchanges because of both being rather sensitivity and weak mutual attraction~\cite{Cui2023}, further leading to persisting violently oscillating average susceptibility (see Fig.~\ref{Sfig:figure1}(a)). In such case, the system does not easily get trapped in a monotonous state such as GC or AP. Moreover, this cluster exhibits much more stable trajectory in comparison with Fig.~1 in Ref\cite{Cui2023} in which the coupled feedback is not considered, showing one advantage of this feedback.

However, intermediate $A$ instead gives rise to GC state, in which the integrated cluster collapses and converges into a single moderate position from which it would never move (see Figs.~\ref{fig:figure1}(b) and (c)), exhibiting sudden drops of $S$ and $SD$ (see Figs.~\ref{fig:figure2}(b) and (c)). Intermediate interaction strength brings about wider opinion spectrum, and thus more insensitive and insistent individuals in presence of the feedback, who can hold more stable position. In addition, at the stage of the opinion cluster, the opinion distances between individuals are mostly within $T$ before the collapse. The above two factors contribute to a global convergence as sharply as possible. It is thus also the evidence for the existence of dynamic equilibrium. Attributing to rather small $\overline{\alpha(t)}$ (see Figs.~\ref{Sfig:figure1}(b) and (c)), individuals become rather slow as the population achieves with a high degree of consensus. Remarkably, we can see in Figs.~\ref{fig:figure3}(b) and (c) that the convergence point is not always the neutral one where all individuals keep the perfect neutrality. The system is definitely initial condition-dependent, which may result from initial unequal distribution of individuals with opinions of different signs. In such case, the convergence direction is uncertain. It is hard to achieve a global convergence towards a neutral consensus when the coupled feedback is present.

As $A$ getting larger, individuals would not keep an open attitude towards distant ones due to strong repulsive forces caused by strong interaction.  The population may get polarized with large $SD$ which is, however, less than $SD_{max}$ (i.e., high-level AP rather than extreme AP). In such case, the population evolves into two opposite camps of different size (see asymmetric opinion distributions illustrated in Figs.~\ref{fig:figure3}(d) and (e)), which thus belongs to asymmetric class~\cite{Pierson2020,Leonard2021}. Opinions of the majority population may firstly self-organize into one integrated cluster to confront the repulsive force, which can leave the moderates within the tolerance range. However, the repulsion from the opposing stubborn extremists caused by the feedback is superior to their mutual attraction, which will reinforce and shift the cluster until the opinion boundaries absorbs it (see Figs.~\ref{fig:figure1}(d) and (e)). As another result, we observe three-step trajectory of $S$ (see Figs.~\ref{fig:figure2}(d) and (e)) and slowly decreasing $\overline{\alpha(t)}$ after an intensive drop (see Figs.~\ref{Sfig:figure1}(d) and (e)), indicating the occurrence of dynamic equilibrium. The system actually enters into a transient stage. The occurrence of dynamic equilibrium have never been uncovered in previous studies, especially those considering the interaction based on bounded confidence. With creasing $A$, high-level polarization would be replaced by symmetric extreme polarization with $SD\approx SD_{max}$ (see Fig.~\ref{fig:figure1}(f), Fig.~\ref{fig:figure2}(f) and Fig.~\ref{fig:figure3}(f)). The cluster can be symmetrically torn apart within a short time (see Fig.~\ref{fig:figure1}(f)), due to the strong repulsive forces arising from strong interaction~\cite{Cui2023}. This makes the prior self-organization into one majority cluster impossible, and individuals be quickly insensitive (see Fig.~\ref{Sfig:figure1}(f) for $\overline{\alpha(t)}\approx0$).

Fig.~\ref{fig:figure4} presents a more clear dependence of $SD$, $S$ and $\chi(S)$ on interaction strength $A$ and tolerance threshold $T$. It is obvious in Fig.~\ref{fig:figure4}(a1) and (a2) that increasing $A$ has an obvious nonlinear effect on both polarization degree $SD$ and opinion diversity i.e., opinion entropy $S$. The range of weak interaction $A<A_{c1}$ is for the emergence of HD state, showing small $SD$ and highland of $S$. We have also checked that spectrum width of the cluster is positively related to the value of $S$. While the system governed by the intermediate range $A_{c1}<A<A_{c2}$ provides sufficient evidence for GC state with $SD=0$ and $S=0$. This is definitely in contrast to the emergent properties generated by the attraction-repulsion model without consideration of the feedback~\cite{Cui2023}, in which a dynamic balance is responsible for the emergence of HD state. Similar behaviors can be observed in Figs.~\ref{fig:figure4}(b1) and (b2). While Figs.~\ref{fig:figure4}(c1) and (c2) reveal that the system undergoes a shift from AP to HD state with increasing tolerance threshold $T$. This is largely in accordance with the empirical evidences~\cite{Spencer2010,Noorazar2020} and the social moral rules requiring a high level of tolerance, essentially attempting to achieve HD state.
In addition, notice that the peaks of $\chi(S)$ corresponds rightly to the critical points between these phases, which further validates the phase-identification method developed in Ref~\cite{Cui2023}.

\begin{figure*}[!ht]
\centering
\includegraphics[width=\linewidth]{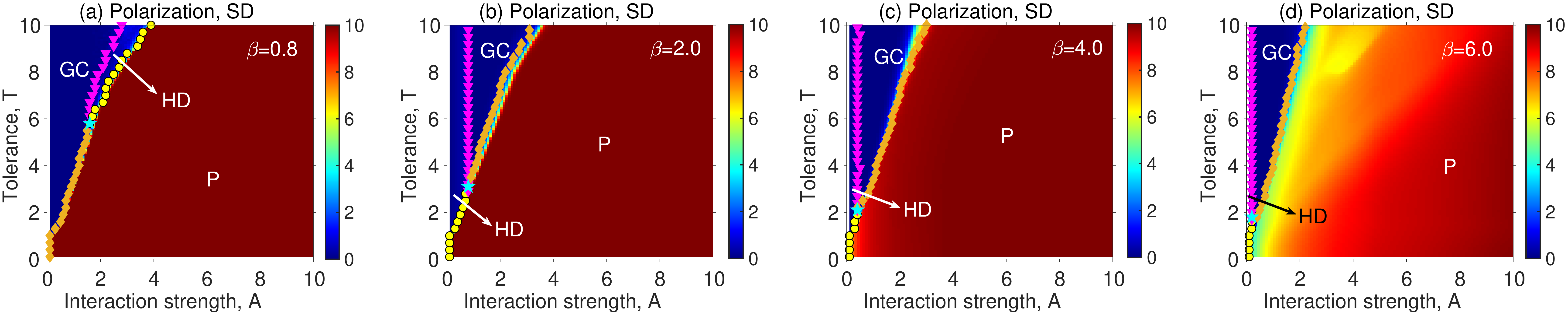}
\caption{Phase diagrams in ($A$,~$T$) space for four different values of $\beta$, showing polarization degree $SD$. The lines consisting of pink triangles separate the regions of HD and GC phases; while the lines consisting of yellow circles depict the boundaries indicated by the second threshold $A_{c2}$, separating the regions of HD and P phases, as given by our method. The lines consisting of brown diamonds separate the regions of GC and P phases. The regions belonging to different states are labeled in the subplots. Especially, the light blue pentagrams illustrated in (b)-(d) indicate the triple points.
}
\label{fig:figure5}
\end{figure*}

We next extend the simulations to $(A,~T)$ space, as shown in Fig.~\ref{fig:figure5}, where the color encodes the values of $SD$ for four different representative values of feedback strength $\beta$. The system finally evolves into either HD or AP state when $\beta$ is small, such that there exists one transition from HD to AP. When $\beta$ increases to some extent ($\beta=2.0$), the intermediate range of $A$ can additionally lead to the emergence of GC state for sufficient tolerance, as well as the triple points (see the blue or black pentagrams illustrated in Figs.~\ref{fig:figure5}(b)-(d)). Therefore, we can observe three different transitions by increasing $A$: from HD to GC which is largely dependent on $A$, from GC to AP and from HD to AP when $\beta$ is larger than a certain value. For all strong feedback promotes GC to erode the regions of HD, and in turn AP occupy the regions of GC, leading to the shrinkage of HD regions. Whatever, both increasing $T$ and decreasing $A$ can prevent the population from being polarized as much as possible for a given $\beta$. However, we find that strong feedback is responsible for the decreasing polarization degree such that extreme AP is widely replaced by high-level AP (see Fig.~\ref{fig:figure5}(d)). Since stubbornness of individuals becomes rather strong (small $\alpha_{i}$) as strengthening interactions bringing about intense struggles between individuals of different signs, which greatly promote the self-organization into one integrated cluster gradually approaching the boundary (see Figs.~\ref{fig:figure1}(d) and (e)), i.e., the dynamic equilibrium. 

\begin{figure*}
\centering
\includegraphics[width=\linewidth]{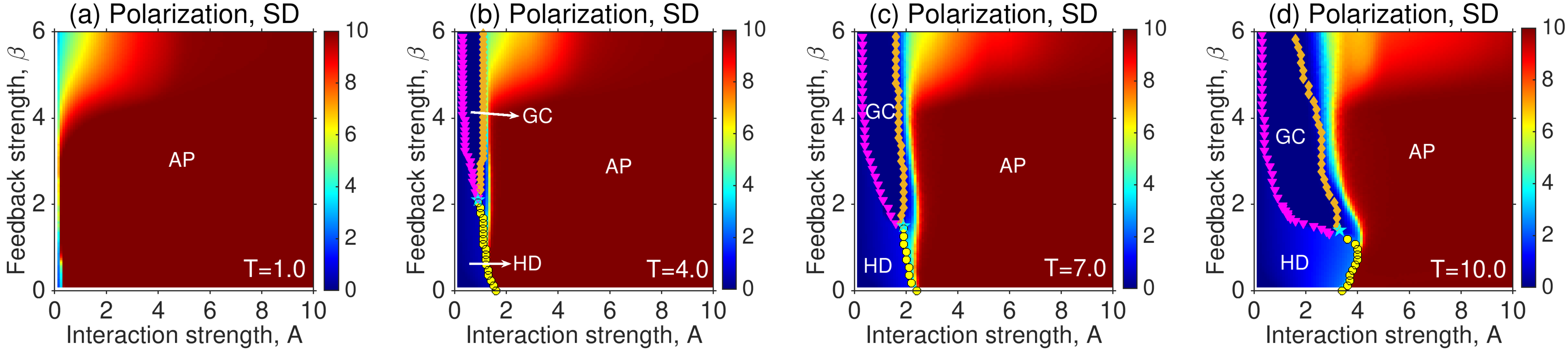}
\caption{Phase diagrams in ($A$,~$\beta$) space for four different values of $T$, showing polarization degree $SD$. The lines consisting of pink triangles separate the regions of GC and HD phases; while the lines consisting of yellow circles depict the boundaries indicated by the second threshold $A_{c2}$, separating the regions of HD and P phases, as given by our method. The lines consisting of brown diamonds separate the regions of GC and P phases. The regions belonging to different states are labeled in the subplots. Especially, the light blue pentagram illustrated in (b) indicates one triple point. The four dots correspond to the subplots of six parameter combinations in Fig.~\ref{fig:figure1}.
}
\label{fig:figure6}
\end{figure*}

Fig.~\ref{fig:figure6} depicts the phase diagrams in ($A$,~$\beta$) space. By increasing $A$, there still occur three transitions when individuals are sufficiently tolerant (see Figs.~\ref{fig:figure6}(b)-(d)): the transition from HD to AP occurs with weak feedback,  whereas strong feedback is favorable for the other two transitions: from GC to AP and from HD to GC. It suggests that interaction strength $A$ plays a decisive role. While larger light red regions of AP (see Fig.~\ref{fig:figure6}) further confirm that strong feedback may prevent the system from getting extremely polarized, and facilitate GC by suppressing HD which, however, becomes dominant for small $\beta$ and $A$, but large $T$.

\begin{figure*}
\centering
\includegraphics[width=\linewidth]{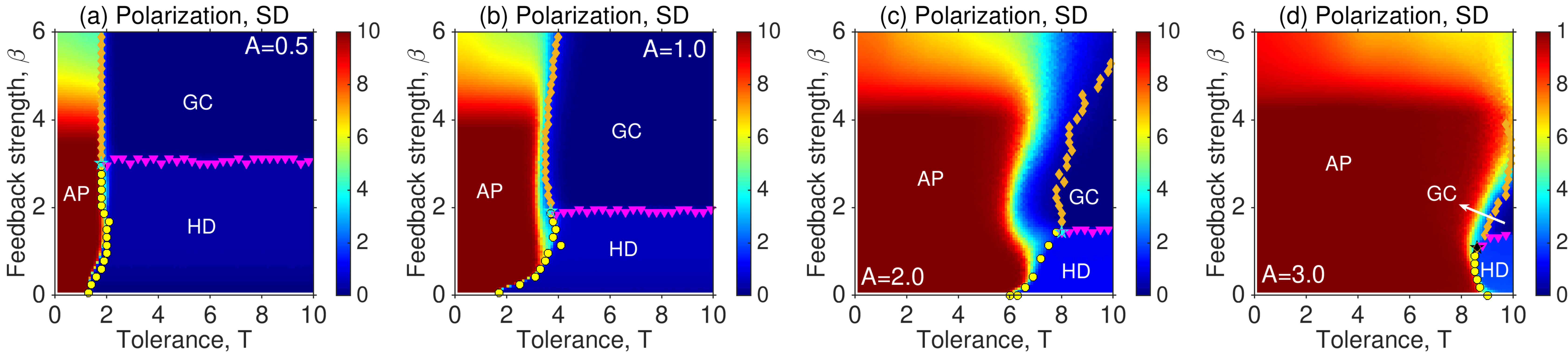}
\caption{Phase diagrams in ($T$,~$\beta$) space for four different values of $A$, showing polarization degree $SD$. The lines consisting of pink triangles separate the regions of GC and HD phases; while the lines consisting of yellow circles depict the boundaries indicated by the second threshold $A_{c2}$, separating the regions of HD and P phases, as given by our method. The lines consisting of brown diamonds separate the regions of GC and P phases. The regions belonging to different states are labeled in the subplots. While the light blue pentagrams indicate the existence of triple points.
}
\label{fig:figure7}
\end{figure*}

Fig.~\ref{fig:figure7} offers a comprehensive view of the effects of individuals' tolerance on opinion dynamics for different levels of feedback. There are not only one transition from HD to GC with increasing $\beta$, but also two novel transitions by increasing $T$: from AP to HD and from AP to GC. We find that, if $A$ is not small, AP is likely to emerge for small $T$, whose regions expands with $A$, along with decreasing likelihood of HD and GC. While increasing both $T$ and $\beta$ is responsible for GC state. In addition, sufficiently tolerant individuals and weak interactions are essential to easily achieving HD state. Still, the system can generate triple points in such parameter space.

Overall, Figs.~\ref{fig:figure5} to \ref{fig:figure7} further confirm that AP may increase with the strengthening feedback, particularly with low tolerance and strong interaction. It supports the researchers' worry that political polarization among the US public may increase due to the AP with the coupled relationship between people's views and their response sensitivity~\cite{Mason2016,Luttig2017,Iyengar2019,Dellaposta2020}. In addition, the phase-identification method allow us to build phase diagrams and to locate where the triple points are, by identifying the boundaries between different phases.        

\begin{figure*}[!ht]
\centering
\includegraphics[width=\linewidth]{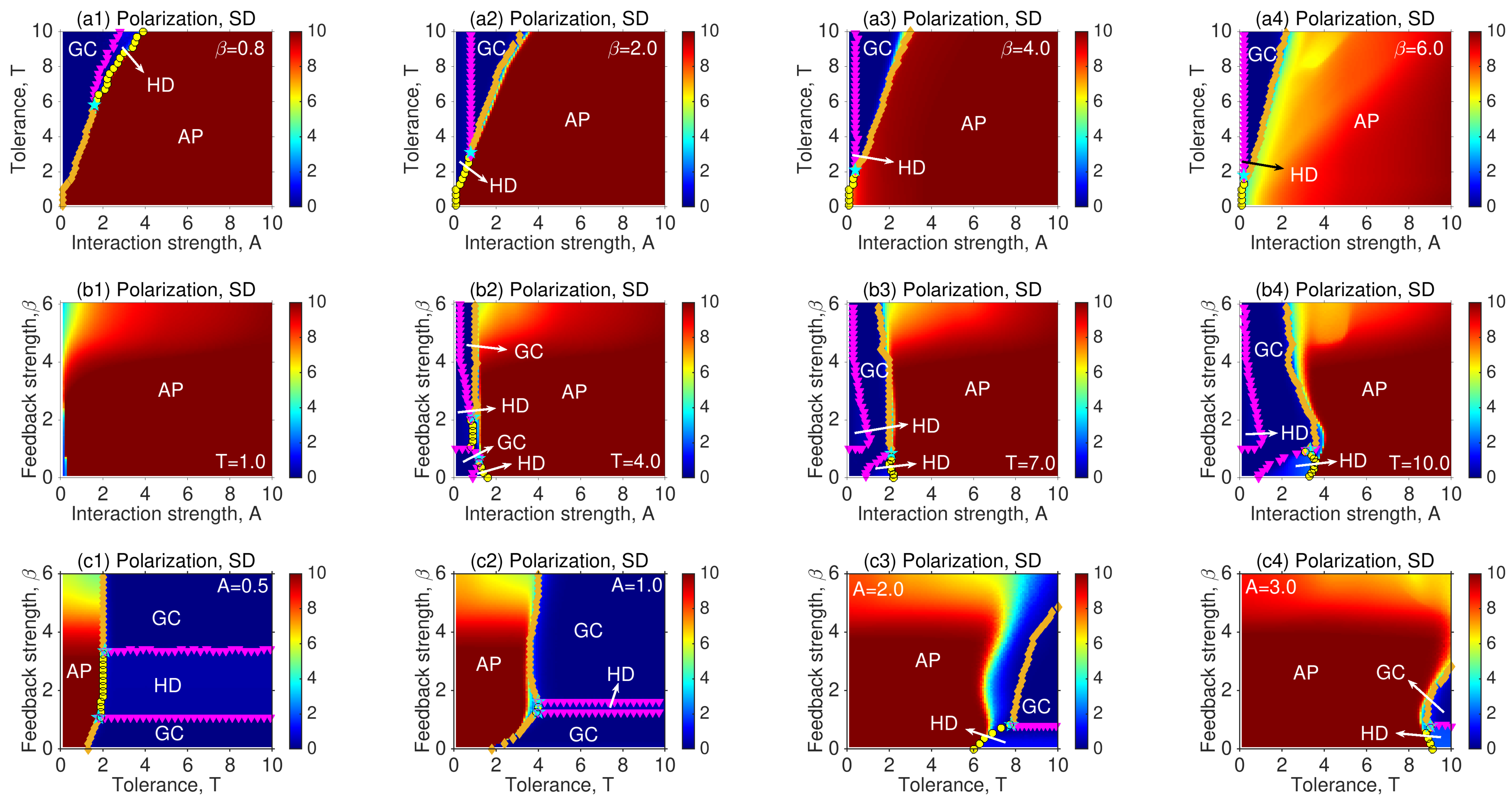}
\caption{
Phase diagram for the population embed on part of Facebook network, where interactions among individuals are fixed. (a1)-(a3) Phase diagrams in $(A,~T)$ space for four different values of $\beta$; (b1)-(b4) Phase diagram in $(A,~\beta)$ space for four different values of $T$; (c1)-(c4) Phase diagram in $(T,~\beta)$ space for four different values of $A$. We run simulations with $SD$ in all subplots. The lines consisting of different markers denote the boundaries between different phases, which are the same as those presented in Fig.~\ref{fig:figure5}. The regions of the three phases are correspondingly labeled in the subplots. The light blue pentagrams indicate the triple points which are more frequent. The degree distribution of the employed network is $p_{k}\sim k^{-\gamma}$ with $\gamma=-3.22$. The other structural parameters are: $N=43952$, mean degree $\langle k\rangle =8.30$, $\langle k^{2}\rangle =205.36$, maximum degree $k_{max}=223$, minimum degree $k_{min}=1$, modularity is $0.59$, clustering coefficient $c=0.12$ and degree correlation coefficient $r\gtrsim 0$. 
}
\label{fig:figure8}
\end{figure*}

Online social networks are increasingly used to access the formation dynamics of opinions with respect to COVID-19 vaccines, gun-control, abortion and so on~\cite{Baumann2020,Wang2020,Johnson2020,Santos2021}. These platforms can reduce barriers and cost to information and, further, allow individuals to freely voice their viewpoints, consequently improving rate of opinion exchanges. We thus subsequently embed our model into the social networks where individuals interact through fixed connections, so as to explore whether the topology of fixed connections can result in the changes of opinion evolution dynamics. Still, similar phase diagrams are presented in Fig.~\ref{fig:figure8} in three different parameter spaces: $(A,~T)$, $(A,~\beta)$ and $(T,\beta)$, suggesting that evolution outcomes are mainly dominated by the model rules, rather than the structure of employed networks. At the same time, a comparison between Figs.~\ref{fig:figure5} to \ref{fig:figure8} reveal that introduction of fixed interactions can remarkably give rise to much larger high-level AP, which is actually attributed to the existence of cluster-level self-reinforced mechanism (see Fig.~\ref{Sfig:figure2})~\cite{Cui2023}. As an another result, both weak feedback and high-level tolerance promote GC regions to erode the regions of HD and to split them into two parts (see Fig.~\ref{fig:figure8}(a1) and (b2)-(b4)). In such case, individuals are generally within the stable clusters and moderate, who are rather sensitive. However, the existence of these clusters can prevent them from contacting and struggling with those of dissimilar opinions, and in turn exert persisting neutralizing influence on the prejudiced ones on the boundaries because of being highly tolerant, regardless of that they are relatively insensitive. Consequently, opinion exchanges are less frequent to maintain more stable opinions, and GC state can be more easily achieved. 

\section{Discussions and Conclusions}
\label{sec:conclusion}
For the first time, this study introduced the feedback to capture the coupled relationship between individuals' opinions and their susceptibilities to the views of others, which has been verified to be a useful approach to the micro foundations of AP dynamics. This assumption allows us to explore the unique dynamic features of AP for the first time. The simulations on both time-varying and static social networks show that strong coupled feedback between individuals' opinions and susceptibility, strong interaction among narrow minded individuals facilitate the emergence of AP state.  Extreme views easily arise from strong interaction between intolerant individuals. Since strong stubbornness is responsible for existence of considerable extremists who can exert persisting influence on the moderate ones in the population. In contrast, weak interaction, high tolerance and weak coupled feedback are favorable for HD state, which actually guide us how to avoid the dilemmas caused by AP. Otherwise, strong stubbornness caused by extreme positions make individuals stand out together to face the opposite of them. In addition, the simulations further confirm that the peak of entropy susceptibility is indeed a sign of transition between GC, HD or AP states~\cite{Cui2023}.

While intermediate ranges of parameters yield GC state which emerges along with that one integrated cluster always collapses and converges towards a moderate position after a long-time persistence, leading to sudden drops of both $S$ and $SD$. It is the first remark. Moreover, the above results highlight the second remark that the negative coupled feedback actually offers a counterforce to establish an inversion between GC and HD, in comparison with the case without this feedback~\cite{Cui2023}. As the third remark, stronger feedback measuring more emotional response, can yield high likelihood of polarization, which is in accordance with empirical investigations of affective polarization in US~\cite{Iyengar2012,Mason2016,Luttig2017,Iyengar2019}. Still, fixed connections can bring about cluster-level self-reinforced mechanism which can enlarge the regions of high-level AP.

Moreover, as the forth remark, the coupled feedback with large intermediate $A$ may promote fast self-organization into one opinion cluster towards GC or even high-level AP. This cluster that does not keep robust against time is also responsible for the step-like dynamic behaviors of opinion entropy. In such case, the system enters into a transient stage, suggesting the existence of dynamic equilibrium. In summary, the above four remarks  cannot be identified by previous studies involving polarization issue, thus novel unique dynamics features of AP within our model framework. 

Our model is based on the simple assumption that interactions between similar actors will reduce their differences, and that the opposite is true for interactions between distant ones. It is significant that the future studies should take into account some empirical characters of individuals which might generate different scenarios, such as heterogeneous duration time of interactions, heterogeneous feedback strength or different social positions. Moreover, the existence of prior self-organization into one integrated cluster before reaching AP state may highlight a golden time window in which regulation measures may most effectively prevent the public from reaching dangerous degrees of AP, which is also worth leaving this investigation to future research. What is more important, empirical evidences to verify the conclusions of the present study is required. Whatever, this study opens one interesting issue to identify the difference between AP and IP with respect their emergent dynamics.

\section*{Acknowledgments}
This work was supported by the Key Program of the National Natural Science Foundation of China (Grant No.~71731002), and by Guangdong Basic and Applied Basic Research Foundation (Grant No.~2021A1515011975).

\section*{Appendix}
\label{sec:appendix}
\begin{figure*}
\centering
\includegraphics[width=\linewidth]{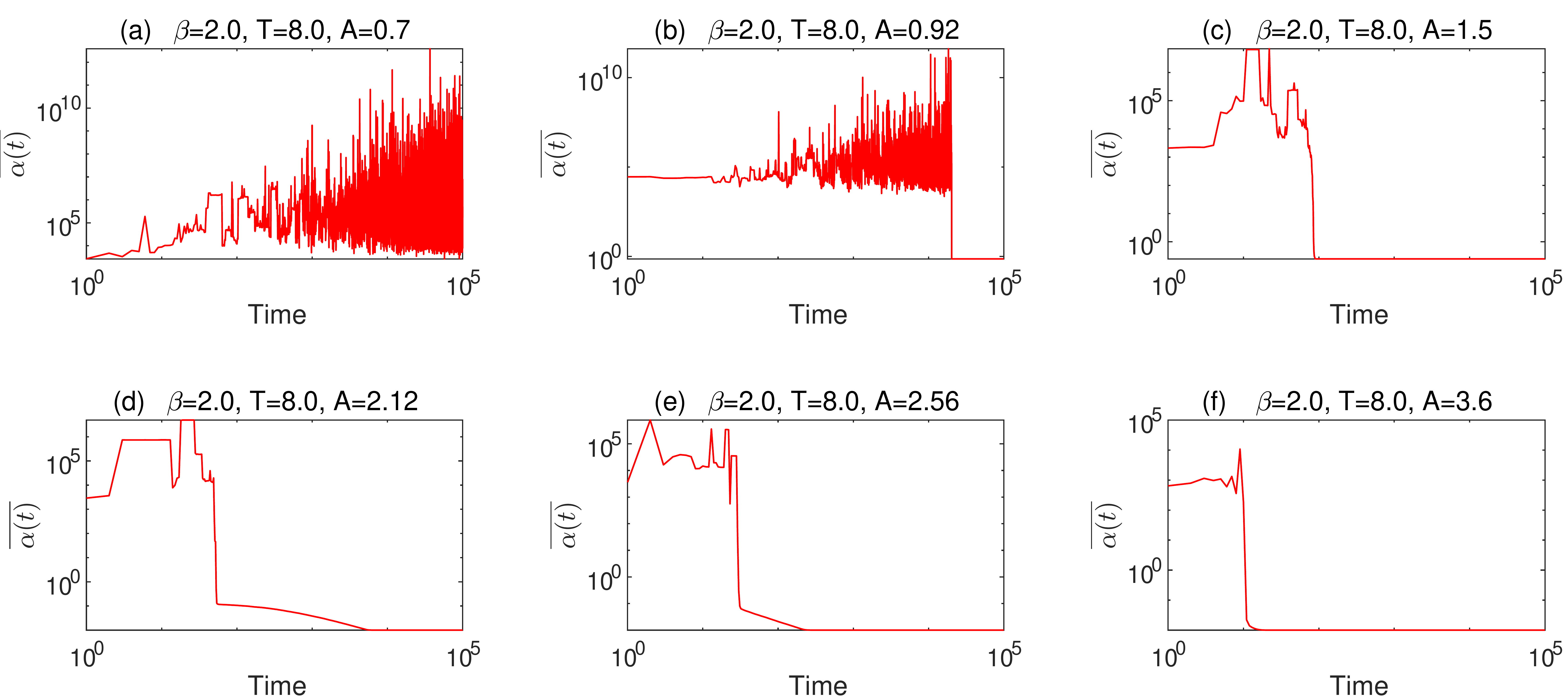}
\caption{
The evolution of average susceptibility of the population, where the values of parameters are listed in the titles of the subfigures, corresponding to the six cases illustrated in Fig.~\ref{fig:figure1}. 
}
\label{Sfig:figure1}
\end{figure*}

\begin{figure*}
\centering
\includegraphics[width=\linewidth]{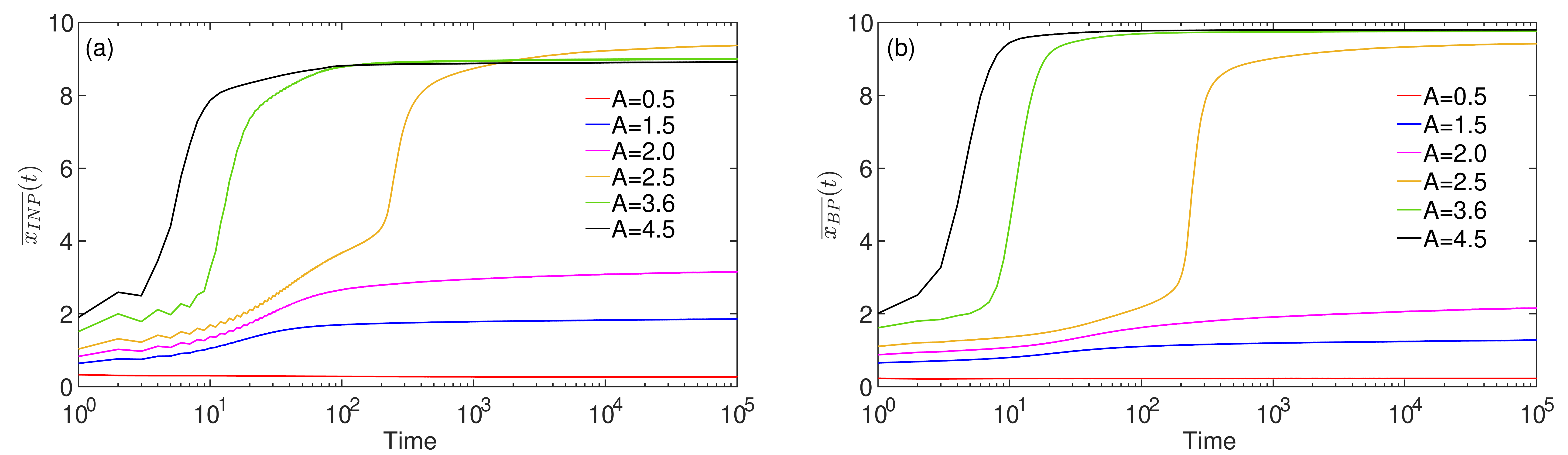}
\caption{
(a) Evolution of average opinion of innermost members owning positive opinions $\overline{x_{INP}}(t)$. (b) Evolution of average opinion of individuals at the boundaries of positive clusters $\overline{x_{BP}}(t)$. The values of other parameters are $T=10.0$ and $\beta=0.5$.
}
\label{Sfig:figure2}
\end{figure*}

Within our model framework, the configurations of positive clusters (i.e. the clusters of positive opinions) are similar to those of negative clusters. We thus report in Fig.~\ref{Sfig:figure2} the evolution of average opinions of the innermost members (who have no neighbours with different signs of opinions) and those at the boundaries of the positive clusters. In the parameter ranges of HD and GC ($A\leq2.5$), the innermost members own much cleared positions than those at the boundaries do, because of dominated mutual attraction caused by weak interaction strength~\cite{Cui2023}. In contrast, as $A$ increasing to the range of AP ($A>2.5$), the individuals at the boundaries of clusters become more extreme than those innermost ones, as one can see in Fig.~\ref{Sfig:figure2} that $\overline{x_{INP}}(t)$ is terminally smaller than $\overline{x_{BP}}(t)$. Cluster-level self-reinforced mechanism forms. In such case, like-minded cluster members persistently support those on the boundaries to repel connected distant ones. Polarization can more easily emerge. On the other hand, one should note that the clustering behaviors prevent the innermost members from being further provoked by those from opposite camps, which is associated with higher likelihood of high-level polarization rather than extreme polarization.   

\bibliography{Reference}

\begin{thebibliography}{46}%
\makeatletter
\providecommand \@ifxundefined [1]{%
 \@ifx{#1\undefined}
}%
\providecommand \@ifnum [1]{%
 \ifnum #1\expandafter \@firstoftwo
 \else \expandafter \@secondoftwo
 \fi
}%
\providecommand \@ifx [1]{%
 \ifx #1\expandafter \@firstoftwo
 \else \expandafter \@secondoftwo
 \fi
}%
\providecommand \natexlab [1]{#1}%
\providecommand \enquote  [1]{``#1''}%
\providecommand \bibnamefont  [1]{#1}%
\providecommand \bibfnamefont [1]{#1}%
\providecommand \citenamefont [1]{#1}%
\providecommand \href@noop [0]{\@secondoftwo}%
\providecommand \href [0]{\begingroup \@sanitize@url \@href}%
\providecommand \@href[1]{\@@startlink{#1}\@@href}%
\providecommand \@@href[1]{\endgroup#1\@@endlink}%
\providecommand \@sanitize@url [0]{\catcode `\\12\catcode `\$12\catcode
  `\&12\catcode `\#12\catcode `\^12\catcode `\_12\catcode `\%12\relax}%
\providecommand \@@startlink[1]{}%
\providecommand \@@endlink[0]{}%
\providecommand \url  [0]{\begingroup\@sanitize@url \@url }%
\providecommand \@url [1]{\endgroup\@href {#1}{\urlprefix }}%
\providecommand \urlprefix  [0]{URL }%
\providecommand \Eprint [0]{\href }%
\providecommand \doibase [0]{http://dx.doi.org/}%
\providecommand \selectlanguage [0]{\@gobble}%
\providecommand \bibinfo  [0]{\@secondoftwo}%
\providecommand \bibfield  [0]{\@secondoftwo}%
\providecommand \translation [1]{[#1]}%
\providecommand \BibitemOpen [0]{}%
\providecommand \bibitemStop [0]{}%
\providecommand \bibitemNoStop [0]{.\EOS\space}%
\providecommand \EOS [0]{\spacefactor3000\relax}%
\providecommand \BibitemShut  [1]{\csname bibitem#1\endcsname}%
\let\auto@bib@innerbib\@empty
\bibitem [{\citenamefont {Axelrod}(1997)}]{Axelrod1997}%
  \BibitemOpen
  \bibfield  {author} {\bibinfo {author} {\bibfnamefont {R.}~\bibnamefont
  {Axelrod}},\ }\href {\doibase 10.1177/0022002797041002001} {\bibfield
  {journal} {\bibinfo  {journal} {Journal of conflict resolution}\ }\textbf
  {\bibinfo {volume} {41}},\ \bibinfo {pages} {203} (\bibinfo {year}
  {1997})}\BibitemShut {NoStop}%
\bibitem [{\citenamefont {Bramson}\ \emph {et~al.}(2017)\citenamefont
  {Bramson}, \citenamefont {Grim}, \citenamefont {Singer}, \citenamefont
  {Berger}, \citenamefont {Sack}, \citenamefont {Fisher}, \citenamefont
  {Flocken},\ and\ \citenamefont {Holman}}]{Bramson2017}%
  \BibitemOpen
  \bibfield  {author} {\bibinfo {author} {\bibfnamefont {A.}~\bibnamefont
  {Bramson}}, \bibinfo {author} {\bibfnamefont {P.}~\bibnamefont {Grim}},
  \bibinfo {author} {\bibfnamefont {D.~J.}\ \bibnamefont {Singer}}, \bibinfo
  {author} {\bibfnamefont {W.~J.}\ \bibnamefont {Berger}}, \bibinfo {author}
  {\bibfnamefont {G.}~\bibnamefont {Sack}}, \bibinfo {author} {\bibfnamefont
  {S.}~\bibnamefont {Fisher}}, \bibinfo {author} {\bibfnamefont
  {C.}~\bibnamefont {Flocken}}, \ and\ \bibinfo {author} {\bibfnamefont
  {B.}~\bibnamefont {Holman}},\ }\href {\doibase
  http://dx.doi.org/10.1086/688938} {\bibfield  {journal} {\bibinfo  {journal}
  {Philosophy of Science}\ }\textbf {\bibinfo {volume} {84}},\ \bibinfo {pages}
  {115} (\bibinfo {year} {2017})}\BibitemShut {NoStop}%
\bibitem [{\citenamefont {Guilbeault}\ \emph {et~al.}(2018)\citenamefont
  {Guilbeault}, \citenamefont {Becker},\ and\ \citenamefont
  {Centola}}]{Guilbeault2018}%
  \BibitemOpen
  \bibfield  {author} {\bibinfo {author} {\bibfnamefont {D.}~\bibnamefont
  {Guilbeault}}, \bibinfo {author} {\bibfnamefont {J.}~\bibnamefont {Becker}},
  \ and\ \bibinfo {author} {\bibfnamefont {D.}~\bibnamefont {Centola}},\ }\href
  {\doibase 10.1073/pnas.172266411} {\bibfield  {journal} {\bibinfo  {journal}
  {Proceedings of the National Academy of Sciences}\ }\textbf {\bibinfo
  {volume} {115}},\ \bibinfo {pages} {9714} (\bibinfo {year}
  {2018})}\BibitemShut {NoStop}%
\bibitem [{\citenamefont {Allcott}\ \emph {et~al.}(2020)\citenamefont
  {Allcott}, \citenamefont {Boxell}, \citenamefont {Conway}, \citenamefont
  {Gentzkow}, \citenamefont {Thaler},\ and\ \citenamefont
  {Yang}}]{Allcott2020}%
  \BibitemOpen
  \bibfield  {author} {\bibinfo {author} {\bibfnamefont {H.}~\bibnamefont
  {Allcott}}, \bibinfo {author} {\bibfnamefont {L.}~\bibnamefont {Boxell}},
  \bibinfo {author} {\bibfnamefont {J.}~\bibnamefont {Conway}}, \bibinfo
  {author} {\bibfnamefont {M.}~\bibnamefont {Gentzkow}}, \bibinfo {author}
  {\bibfnamefont {M.}~\bibnamefont {Thaler}}, \ and\ \bibinfo {author}
  {\bibfnamefont {D.}~\bibnamefont {Yang}},\ }\href {\doibase
  10.1016/j.jpubeco.2020.104254} {\bibfield  {journal} {\bibinfo  {journal}
  {Journal of public economics}\ }\textbf {\bibinfo {volume} {191}},\ \bibinfo
  {pages} {104254} (\bibinfo {year} {2020})}\BibitemShut {NoStop}%
\bibitem [{\citenamefont {Castellano}\ \emph {et~al.}(2009)\citenamefont
  {Castellano}, \citenamefont {Fortunato},\ and\ \citenamefont
  {Loreto}}]{Castellano2009}%
  \BibitemOpen
  \bibfield  {author} {\bibinfo {author} {\bibfnamefont {C.}~\bibnamefont
  {Castellano}}, \bibinfo {author} {\bibfnamefont {S.}~\bibnamefont
  {Fortunato}}, \ and\ \bibinfo {author} {\bibfnamefont {V.}~\bibnamefont
  {Loreto}},\ }\href {\doibase 10.1103/RevModPhys.81.591} {\bibfield  {journal}
  {\bibinfo  {journal} {Reviews of modern physics}\ }\textbf {\bibinfo {volume}
  {81}},\ \bibinfo {pages} {591} (\bibinfo {year} {2009})}\BibitemShut
  {NoStop}%
\bibitem [{\citenamefont {Sobkowicz}\ \emph {et~al.}(2012)\citenamefont
  {Sobkowicz}, \citenamefont {Kaschesky},\ and\ \citenamefont
  {Bouchard}}]{Sobkowicz2012}%
  \BibitemOpen
  \bibfield  {author} {\bibinfo {author} {\bibfnamefont {P.}~\bibnamefont
  {Sobkowicz}}, \bibinfo {author} {\bibfnamefont {M.}~\bibnamefont
  {Kaschesky}}, \ and\ \bibinfo {author} {\bibfnamefont {G.}~\bibnamefont
  {Bouchard}},\ }\href {\doibase 10.1016/j.giq.2012.06.005} {\bibfield
  {journal} {\bibinfo  {journal} {Government Information Quarterly}\ }\textbf
  {\bibinfo {volume} {29}},\ \bibinfo {pages} {470} (\bibinfo {year}
  {2012})}\BibitemShut {NoStop}%
\bibitem [{\citenamefont {Moe}\ and\ \citenamefont
  {Schweidel}(2012)}]{Moe2012}%
  \BibitemOpen
  \bibfield  {author} {\bibinfo {author} {\bibfnamefont {W.~W.}\ \bibnamefont
  {Moe}}\ and\ \bibinfo {author} {\bibfnamefont {D.~A.}\ \bibnamefont
  {Schweidel}},\ }\href {\doibase 10.1287/mksc.1110.0662} {\bibfield  {journal}
  {\bibinfo  {journal} {Marketing Science}\ }\textbf {\bibinfo {volume} {31}},\
  \bibinfo {pages} {372} (\bibinfo {year} {2012})}\BibitemShut {NoStop}%
\bibitem [{\citenamefont {Lelkes}(2016)}]{Lelkes2016}%
  \BibitemOpen
  \bibfield  {author} {\bibinfo {author} {\bibfnamefont {Y.}~\bibnamefont
  {Lelkes}},\ }\href {\doibase 10.1093/poq/nfw005} {\bibfield  {journal}
  {\bibinfo  {journal} {Public Opinion Quarterly}\ }\textbf {\bibinfo {volume}
  {80}},\ \bibinfo {pages} {392} (\bibinfo {year} {2016})}\BibitemShut
  {NoStop}%
\bibitem [{\citenamefont {Iyengar}\ \emph {et~al.}(2012)\citenamefont
  {Iyengar}, \citenamefont {Sood},\ and\ \citenamefont {Lelkes}}]{Iyengar2012}%
  \BibitemOpen
  \bibfield  {author} {\bibinfo {author} {\bibfnamefont {S.}~\bibnamefont
  {Iyengar}}, \bibinfo {author} {\bibfnamefont {G.}~\bibnamefont {Sood}}, \
  and\ \bibinfo {author} {\bibfnamefont {Y.}~\bibnamefont {Lelkes}},\ }\href
  {\doibase 10.1093/poq/nfs038} {\bibfield  {journal} {\bibinfo  {journal}
  {Public opinion quarterly}\ }\textbf {\bibinfo {volume} {76}},\ \bibinfo
  {pages} {405} (\bibinfo {year} {2012})}\BibitemShut {NoStop}%
\bibitem [{\citenamefont {Lauka}\ \emph {et~al.}(2018)\citenamefont {Lauka},
  \citenamefont {McCoy},\ and\ \citenamefont {Firat}}]{Lauka2018}%
  \BibitemOpen
  \bibfield  {author} {\bibinfo {author} {\bibfnamefont {A.}~\bibnamefont
  {Lauka}}, \bibinfo {author} {\bibfnamefont {J.}~\bibnamefont {McCoy}}, \ and\
  \bibinfo {author} {\bibfnamefont {R.~B.}\ \bibnamefont {Firat}},\ }\href
  {\doibase 10.1177/0002764218759581} {\bibfield  {journal} {\bibinfo
  {journal} {American behavioral scientist}\ }\textbf {\bibinfo {volume}
  {62}},\ \bibinfo {pages} {107} (\bibinfo {year} {2018})}\BibitemShut
  {NoStop}%
\bibitem [{\citenamefont {Iyengar}\ \emph {et~al.}(2019)\citenamefont
  {Iyengar}, \citenamefont {Lelkes}, \citenamefont {Levendusky}, \citenamefont
  {Malhotra},\ and\ \citenamefont {Westwood}}]{Iyengar2019}%
  \BibitemOpen
  \bibfield  {author} {\bibinfo {author} {\bibfnamefont {S.}~\bibnamefont
  {Iyengar}}, \bibinfo {author} {\bibfnamefont {Y.}~\bibnamefont {Lelkes}},
  \bibinfo {author} {\bibfnamefont {M.}~\bibnamefont {Levendusky}}, \bibinfo
  {author} {\bibfnamefont {N.}~\bibnamefont {Malhotra}}, \ and\ \bibinfo
  {author} {\bibfnamefont {S.~J.}\ \bibnamefont {Westwood}},\ }\href {\doibase
  10.1146/annurev-polisci-051117-073034} {\bibfield  {journal} {\bibinfo
  {journal} {Annual Review of Political Science}\ }\textbf {\bibinfo {volume}
  {22}},\ \bibinfo {pages} {129} (\bibinfo {year} {2019})}\BibitemShut
  {NoStop}%
\bibitem [{\citenamefont {Boxell}\ \emph {et~al.}(2020)\citenamefont {Boxell},
  \citenamefont {Gentzkow},\ and\ \citenamefont {Shapiro}}]{Boxell2020}%
  \BibitemOpen
  \bibfield  {author} {\bibinfo {author} {\bibfnamefont {L.}~\bibnamefont
  {Boxell}}, \bibinfo {author} {\bibfnamefont {M.}~\bibnamefont {Gentzkow}}, \
  and\ \bibinfo {author} {\bibfnamefont {J.~M.}\ \bibnamefont {Shapiro}},\
  }\href {\doibase 10.1162/rest_a_01160} {\bibfield  {journal} {\bibinfo
  {journal} {The Review of Economics and Statistics}\ ,\ \bibinfo {pages} {1}}
  (\bibinfo {year} {2020})}\BibitemShut {NoStop}%
\bibitem [{\citenamefont {Reiljan}(2020)}]{Reiljan2020}%
  \BibitemOpen
  \bibfield  {author} {\bibinfo {author} {\bibfnamefont {A.}~\bibnamefont
  {Reiljan}},\ }\href {\doibase 10.1111/1475-6765.12351} {\bibfield  {journal}
  {\bibinfo  {journal} {European journal of political research}\ }\textbf
  {\bibinfo {volume} {59}},\ \bibinfo {pages} {376} (\bibinfo {year}
  {2020})}\BibitemShut {NoStop}%
\bibitem [{\citenamefont {Wagner}(2021)}]{Wagner2021}%
  \BibitemOpen
  \bibfield  {author} {\bibinfo {author} {\bibfnamefont {M.}~\bibnamefont
  {Wagner}},\ }\href {\doibase 10.1016/j.electstud.2020.102199} {\bibfield
  {journal} {\bibinfo  {journal} {Electoral Studies}\ }\textbf {\bibinfo
  {volume} {69}},\ \bibinfo {pages} {102199} (\bibinfo {year}
  {2021})}\BibitemShut {NoStop}%
\bibitem [{\citenamefont {Abramowitz}\ and\ \citenamefont
  {Saunders}(2008)}]{Abramowitz2008}%
  \BibitemOpen
  \bibfield  {author} {\bibinfo {author} {\bibfnamefont {A.~I.}\ \bibnamefont
  {Abramowitz}}\ and\ \bibinfo {author} {\bibfnamefont {K.~L.}\ \bibnamefont
  {Saunders}},\ }\href {\doibase 10.1017/S0022381608080493} {\bibfield
  {journal} {\bibinfo  {journal} {The Journal of Politics}\ }\textbf {\bibinfo
  {volume} {70}},\ \bibinfo {pages} {542} (\bibinfo {year} {2008})}\BibitemShut
  {NoStop}%
\bibitem [{\citenamefont {Axelrod}\ \emph {et~al.}(2021)\citenamefont
  {Axelrod}, \citenamefont {Daymude},\ and\ \citenamefont
  {Forrest}}]{Axelrod2021}%
  \BibitemOpen
  \bibfield  {author} {\bibinfo {author} {\bibfnamefont {R.}~\bibnamefont
  {Axelrod}}, \bibinfo {author} {\bibfnamefont {J.~J.}\ \bibnamefont
  {Daymude}}, \ and\ \bibinfo {author} {\bibfnamefont {S.}~\bibnamefont
  {Forrest}},\ }\href {\doibase 10.1073/pnas.2102139118} {\bibfield  {journal}
  {\bibinfo  {journal} {Proceedings of the National Academy of Sciences}\
  }\textbf {\bibinfo {volume} {118}},\ \bibinfo {pages} {e2102139118} (\bibinfo
  {year} {2021})}\BibitemShut {NoStop}%
\bibitem [{\citenamefont {Grossman}\ and\ \citenamefont
  {Helpman}(2021)}]{Grossman2021}%
  \BibitemOpen
  \bibfield  {author} {\bibinfo {author} {\bibfnamefont {G.~M.}\ \bibnamefont
  {Grossman}}\ and\ \bibinfo {author} {\bibfnamefont {E.}~\bibnamefont
  {Helpman}},\ }\href {\doibase 10.1093/restud/rdaa031} {\bibfield  {journal}
  {\bibinfo  {journal} {The Review of Economic Studies}\ }\textbf {\bibinfo
  {volume} {88}},\ \bibinfo {pages} {1101} (\bibinfo {year}
  {2021})}\BibitemShut {NoStop}%
\bibitem [{\citenamefont {Perrings}\ \emph {et~al.}(2021)\citenamefont
  {Perrings}, \citenamefont {Hechter},\ and\ \citenamefont
  {Mamada}}]{Perrings2021}%
  \BibitemOpen
  \bibfield  {author} {\bibinfo {author} {\bibfnamefont {C.}~\bibnamefont
  {Perrings}}, \bibinfo {author} {\bibfnamefont {M.}~\bibnamefont {Hechter}}, \
  and\ \bibinfo {author} {\bibfnamefont {R.}~\bibnamefont {Mamada}},\ }\href
  {\doibase 10.1073/pnas.2102145118} {\bibfield  {journal} {\bibinfo  {journal}
  {Proceedings of the National Academy of Sciences}\ }\textbf {\bibinfo
  {volume} {118}},\ \bibinfo {pages} {e2102145118} (\bibinfo {year}
  {2021})}\BibitemShut {NoStop}%
\bibitem [{\citenamefont {Vasconcelos}\ \emph {et~al.}(2021)\citenamefont
  {Vasconcelos}, \citenamefont {Constantino}, \citenamefont {Dannenberg},
  \citenamefont {Lumkowsky}, \citenamefont {Weber},\ and\ \citenamefont
  {Levin}}]{Vasconcelos2021}%
  \BibitemOpen
  \bibfield  {author} {\bibinfo {author} {\bibfnamefont {V.~V.}\ \bibnamefont
  {Vasconcelos}}, \bibinfo {author} {\bibfnamefont {S.~M.}\ \bibnamefont
  {Constantino}}, \bibinfo {author} {\bibfnamefont {A.}~\bibnamefont
  {Dannenberg}}, \bibinfo {author} {\bibfnamefont {M.}~\bibnamefont
  {Lumkowsky}}, \bibinfo {author} {\bibfnamefont {E.}~\bibnamefont {Weber}}, \
  and\ \bibinfo {author} {\bibfnamefont {S.}~\bibnamefont {Levin}},\ }\href
  {\doibase 10.1073/pnas.2102153118} {\bibfield  {journal} {\bibinfo  {journal}
  {Proceedings of the National Academy of Sciences}\ }\textbf {\bibinfo
  {volume} {118}},\ \bibinfo {pages} {e2102153118} (\bibinfo {year}
  {2021})}\BibitemShut {NoStop}%
\bibitem [{\citenamefont {Lelkes}\ and\ \citenamefont
  {Westwood}(2017)}]{Lelkes2017}%
  \BibitemOpen
  \bibfield  {author} {\bibinfo {author} {\bibfnamefont {Y.}~\bibnamefont
  {Lelkes}}\ and\ \bibinfo {author} {\bibfnamefont {S.~J.}\ \bibnamefont
  {Westwood}},\ }\href {\doibase 10.1086/688223} {\bibfield  {journal}
  {\bibinfo  {journal} {The Journal of Politics}\ }\textbf {\bibinfo {volume}
  {79}},\ \bibinfo {pages} {485} (\bibinfo {year} {2017})}\BibitemShut
  {NoStop}%
\bibitem [{\citenamefont {Woolhandler}\ \emph {et~al.}(2021)\citenamefont
  {Woolhandler}, \citenamefont {Himmelstein}, \citenamefont {Ahmed},
  \citenamefont {Bailey}, \citenamefont {Bassett}, \citenamefont {Bird},
  \citenamefont {Bor}, \citenamefont {Bor}, \citenamefont {Carrasquillo},
  \citenamefont {Chowkwanyun} \emph {et~al.}}]{Woolhandler2021}%
  \BibitemOpen
  \bibfield  {author} {\bibinfo {author} {\bibfnamefont {S.}~\bibnamefont
  {Woolhandler}}, \bibinfo {author} {\bibfnamefont {D.~U.}\ \bibnamefont
  {Himmelstein}}, \bibinfo {author} {\bibfnamefont {S.}~\bibnamefont {Ahmed}},
  \bibinfo {author} {\bibfnamefont {Z.}~\bibnamefont {Bailey}}, \bibinfo
  {author} {\bibfnamefont {M.~T.}\ \bibnamefont {Bassett}}, \bibinfo {author}
  {\bibfnamefont {M.}~\bibnamefont {Bird}}, \bibinfo {author} {\bibfnamefont
  {J.}~\bibnamefont {Bor}}, \bibinfo {author} {\bibfnamefont {D.}~\bibnamefont
  {Bor}}, \bibinfo {author} {\bibfnamefont {O.}~\bibnamefont {Carrasquillo}},
  \bibinfo {author} {\bibfnamefont {M.}~\bibnamefont {Chowkwanyun}},  \emph
  {et~al.},\ }\href {\doibase 10.2139/ssrn.199668} {\bibfield  {journal}
  {\bibinfo  {journal} {The Lancet}\ }\textbf {\bibinfo {volume} {397}},\
  \bibinfo {pages} {705} (\bibinfo {year} {2021})}\BibitemShut {NoStop}%
\bibitem [{\citenamefont {Sunstein}(1999)}]{Sunstein1999}%
  \BibitemOpen
  \bibfield  {author} {\bibinfo {author} {\bibfnamefont {C.~R.}\ \bibnamefont
  {Sunstein}},\ }\href {\doibase 10.2139/ssrn.199668} {\bibfield  {journal}
  {\bibinfo  {journal} {University of Chicago Law School, John M. Olin Law \&
  Economics Working Paper}\ } (\bibinfo {year} {1999}),\
  10.2139/ssrn.199668}\BibitemShut {NoStop}%
\bibitem [{\citenamefont {DiMaggio}\ \emph {et~al.}(1996)\citenamefont
  {DiMaggio}, \citenamefont {Evans},\ and\ \citenamefont
  {Bryson}}]{Dimaggio1996}%
  \BibitemOpen
  \bibfield  {author} {\bibinfo {author} {\bibfnamefont {P.}~\bibnamefont
  {DiMaggio}}, \bibinfo {author} {\bibfnamefont {J.}~\bibnamefont {Evans}}, \
  and\ \bibinfo {author} {\bibfnamefont {B.}~\bibnamefont {Bryson}},\ }\href
  {\doibase 10.1086/230995} {\bibfield  {journal} {\bibinfo  {journal}
  {American journal of Sociology}\ }\textbf {\bibinfo {volume} {102}},\
  \bibinfo {pages} {690} (\bibinfo {year} {1996})}\BibitemShut {NoStop}%
\bibitem [{\citenamefont {Kossinets}\ and\ \citenamefont
  {Watts}(2009)}]{Kossinets2009}%
  \BibitemOpen
  \bibfield  {author} {\bibinfo {author} {\bibfnamefont {G.}~\bibnamefont
  {Kossinets}}\ and\ \bibinfo {author} {\bibfnamefont {D.~J.}\ \bibnamefont
  {Watts}},\ }\href {\doibase 10.1086/599247} {\bibfield  {journal} {\bibinfo
  {journal} {American journal of sociology}\ }\textbf {\bibinfo {volume}
  {115}},\ \bibinfo {pages} {405} (\bibinfo {year} {2009})}\BibitemShut
  {NoStop}%
\bibitem [{\citenamefont {Vasconcelos}\ \emph {et~al.}(2019)\citenamefont
  {Vasconcelos}, \citenamefont {Levin},\ and\ \citenamefont
  {Pinheiro}}]{Vasconcelos2019}%
  \BibitemOpen
  \bibfield  {author} {\bibinfo {author} {\bibfnamefont {V.~V.}\ \bibnamefont
  {Vasconcelos}}, \bibinfo {author} {\bibfnamefont {S.~A.}\ \bibnamefont
  {Levin}}, \ and\ \bibinfo {author} {\bibfnamefont {F.~L.}\ \bibnamefont
  {Pinheiro}},\ }\href {\doibase 10.1098/rsif.2019.0196} {\bibfield  {journal}
  {\bibinfo  {journal} {Journal of the Royal Society Interface}\ }\textbf
  {\bibinfo {volume} {16}},\ \bibinfo {pages} {20190196} (\bibinfo {year}
  {2019})}\BibitemShut {NoStop}%
\bibitem [{\citenamefont {Baumann}\ \emph {et~al.}(2020)\citenamefont
  {Baumann}, \citenamefont {Lorenz-Spreen}, \citenamefont {Sokolov},\ and\
  \citenamefont {Starnini}}]{Baumann2020}%
  \BibitemOpen
  \bibfield  {author} {\bibinfo {author} {\bibfnamefont {F.}~\bibnamefont
  {Baumann}}, \bibinfo {author} {\bibfnamefont {P.}~\bibnamefont
  {Lorenz-Spreen}}, \bibinfo {author} {\bibfnamefont {I.~M.}\ \bibnamefont
  {Sokolov}}, \ and\ \bibinfo {author} {\bibfnamefont {M.}~\bibnamefont
  {Starnini}},\ }\href {\doibase 10.1103/PhysRevLett.124.048301} {\bibfield
  {journal} {\bibinfo  {journal} {Physical Review Letters}\ }\textbf {\bibinfo
  {volume} {124}},\ \bibinfo {pages} {048301} (\bibinfo {year}
  {2020})}\BibitemShut {NoStop}%
\bibitem [{\citenamefont {Santos}\ \emph {et~al.}(2021)\citenamefont {Santos},
  \citenamefont {Lelkes},\ and\ \citenamefont {Levin}}]{Santos2021}%
  \BibitemOpen
  \bibfield  {author} {\bibinfo {author} {\bibfnamefont {F.~P.}\ \bibnamefont
  {Santos}}, \bibinfo {author} {\bibfnamefont {Y.}~\bibnamefont {Lelkes}}, \
  and\ \bibinfo {author} {\bibfnamefont {S.~A.}\ \bibnamefont {Levin}},\ }\href
  {\doibase 10.1073/pnas.2102141118} {\bibfield  {journal} {\bibinfo  {journal}
  {Proceedings of the National Academy of Sciences}\ }\textbf {\bibinfo
  {volume} {118}},\ \bibinfo {pages} {e2102141118} (\bibinfo {year}
  {2021})}\BibitemShut {NoStop}%
\bibitem [{\citenamefont {Chu}\ \emph {et~al.}(2021)\citenamefont {Chu},
  \citenamefont {Donges}, \citenamefont {Robertson},\ and\ \citenamefont
  {Pop-Eleches}}]{Chu2021}%
  \BibitemOpen
  \bibfield  {author} {\bibinfo {author} {\bibfnamefont {O.~J.}\ \bibnamefont
  {Chu}}, \bibinfo {author} {\bibfnamefont {J.~F.}\ \bibnamefont {Donges}},
  \bibinfo {author} {\bibfnamefont {G.~B.}\ \bibnamefont {Robertson}}, \ and\
  \bibinfo {author} {\bibfnamefont {G.}~\bibnamefont {Pop-Eleches}},\ }\href
  {\doibase 10.1073/pnas.2104194118} {\bibfield  {journal} {\bibinfo  {journal}
  {Proceedings of the National Academy of Sciences}\ }\textbf {\bibinfo
  {volume} {118}},\ \bibinfo {pages} {e2104194118} (\bibinfo {year}
  {2021})}\BibitemShut {NoStop}%
\bibitem [{\citenamefont {Jusup}\ \emph {et~al.}(2022)\citenamefont {Jusup},
  \citenamefont {Holme}, \citenamefont {Kanazawa}, \citenamefont {Takayasu},
  \citenamefont {Romi{\'c}}, \citenamefont {Wang}, \citenamefont {Ge{\v{c}}ek},
  \citenamefont {Lipi{\'c}}, \citenamefont {Podobnik}, \citenamefont {Wang}
  \emph {et~al.}}]{Jusup2022}%
  \BibitemOpen
  \bibfield  {author} {\bibinfo {author} {\bibfnamefont {M.}~\bibnamefont
  {Jusup}}, \bibinfo {author} {\bibfnamefont {P.}~\bibnamefont {Holme}},
  \bibinfo {author} {\bibfnamefont {K.}~\bibnamefont {Kanazawa}}, \bibinfo
  {author} {\bibfnamefont {M.}~\bibnamefont {Takayasu}}, \bibinfo {author}
  {\bibfnamefont {I.}~\bibnamefont {Romi{\'c}}}, \bibinfo {author}
  {\bibfnamefont {Z.}~\bibnamefont {Wang}}, \bibinfo {author} {\bibfnamefont
  {S.}~\bibnamefont {Ge{\v{c}}ek}}, \bibinfo {author} {\bibfnamefont
  {T.}~\bibnamefont {Lipi{\'c}}}, \bibinfo {author} {\bibfnamefont
  {B.}~\bibnamefont {Podobnik}}, \bibinfo {author} {\bibfnamefont
  {L.}~\bibnamefont {Wang}},  \emph {et~al.},\ }\href {\doibase
  10.1016/j.physrep.2021.10.005} {\bibfield  {journal} {\bibinfo  {journal}
  {Physics Reports}\ }\textbf {\bibinfo {volume} {948}},\ \bibinfo {pages} {1}
  (\bibinfo {year} {2022})}\BibitemShut {NoStop}%
\bibitem [{\citenamefont {Mason}(2016)}]{Mason2016}%
  \BibitemOpen
  \bibfield  {author} {\bibinfo {author} {\bibfnamefont {L.}~\bibnamefont
  {Mason}},\ }\href {\doibase 10.1093/poq/nfw001} {\bibfield  {journal}
  {\bibinfo  {journal} {Public Opinion Quarterly}\ }\textbf {\bibinfo {volume}
  {80}},\ \bibinfo {pages} {351} (\bibinfo {year} {2016})}\BibitemShut
  {NoStop}%
\bibitem [{\citenamefont {Luttig}\ \emph {et~al.}(2017)\citenamefont {Luttig},
  \citenamefont {Federico},\ and\ \citenamefont {Lavine}}]{Luttig2017}%
  \BibitemOpen
  \bibfield  {author} {\bibinfo {author} {\bibfnamefont {M.~D.}\ \bibnamefont
  {Luttig}}, \bibinfo {author} {\bibfnamefont {C.~M.}\ \bibnamefont
  {Federico}}, \ and\ \bibinfo {author} {\bibfnamefont {H.}~\bibnamefont
  {Lavine}},\ }\href {\doibase 10.1177/2053168017737411} {\bibfield  {journal}
  {\bibinfo  {journal} {Research \& politics}\ }\textbf {\bibinfo {volume}
  {4}},\ \bibinfo {pages} {2053168017737411} (\bibinfo {year}
  {2017})}\BibitemShut {NoStop}%
\bibitem [{\citenamefont {DellaPosta}(2020)}]{Dellaposta2020}%
  \BibitemOpen
  \bibfield  {author} {\bibinfo {author} {\bibfnamefont {D.}~\bibnamefont
  {DellaPosta}},\ }\href {\doibase 10.1177/0003122420922989} {\bibfield
  {journal} {\bibinfo  {journal} {American Sociological Review}\ }\textbf
  {\bibinfo {volume} {85}},\ \bibinfo {pages} {507} (\bibinfo {year}
  {2020})}\BibitemShut {NoStop}%
\bibitem [{\citenamefont {Schaffner}\ \emph {et~al.}(2016)\citenamefont
  {Schaffner}, \citenamefont {MacWilliams},\ and\ \citenamefont
  {Nteta}}]{Schaffner2016}%
  \BibitemOpen
  \bibfield  {author} {\bibinfo {author} {\bibfnamefont {B.~F.}\ \bibnamefont
  {Schaffner}}, \bibinfo {author} {\bibfnamefont {M.}~\bibnamefont
  {MacWilliams}}, \ and\ \bibinfo {author} {\bibfnamefont {T.}~\bibnamefont
  {Nteta}},\ }in\ \href {\doibase 10.1002/polq.12737} {\emph {\bibinfo
  {booktitle} {Conference on the US Elections of}}}\ (\bibinfo {year} {2016})\
  pp.\ \bibinfo {pages} {8--9}\BibitemShut {NoStop}%
\bibitem [{\citenamefont {Sides}\ \emph {et~al.}(2017)\citenamefont {Sides},
  \citenamefont {Tesler},\ and\ \citenamefont {Vavreck}}]{Sides2016}%
  \BibitemOpen
  \bibfield  {author} {\bibinfo {author} {\bibfnamefont {J.}~\bibnamefont
  {Sides}}, \bibinfo {author} {\bibfnamefont {M.}~\bibnamefont {Tesler}}, \
  and\ \bibinfo {author} {\bibfnamefont {L.}~\bibnamefont {Vavreck}},\ }\href
  {\doibase 10.1353/jod.2017.0022} {\bibfield  {journal} {\bibinfo  {journal}
  {Journal of Democracy}\ }\textbf {\bibinfo {volume} {28}},\ \bibinfo {pages}
  {34} (\bibinfo {year} {2017})}\BibitemShut {NoStop}%
\bibitem [{\citenamefont {Mitrea}\ \emph {et~al.}(2021)\citenamefont {Mitrea},
  \citenamefont {M{\"u}hlb{\"o}ck},\ and\ \citenamefont
  {Warmuth}}]{Mitrea2021}%
  \BibitemOpen
  \bibfield  {author} {\bibinfo {author} {\bibfnamefont {E.~C.}\ \bibnamefont
  {Mitrea}}, \bibinfo {author} {\bibfnamefont {M.}~\bibnamefont
  {M{\"u}hlb{\"o}ck}}, \ and\ \bibinfo {author} {\bibfnamefont
  {J.}~\bibnamefont {Warmuth}},\ }\href {\doibase 10.1007/s11109-020-09593-7}
  {\bibfield  {journal} {\bibinfo  {journal} {Political Behavior}\ }\textbf
  {\bibinfo {volume} {43}},\ \bibinfo {pages} {785} (\bibinfo {year}
  {2021})}\BibitemShut {NoStop}%
\bibitem [{\citenamefont {Cui}(2023)}]{Cui2023}%
  \BibitemOpen
  \bibfield  {author} {\bibinfo {author} {\bibfnamefont {P.-B.}\ \bibnamefont
  {Cui}},\ }\href {\doibase 10.1016/j.physa.2023.128714} {\bibfield  {journal}
  {\bibinfo  {journal} {Physica A: Statistical Mechanics and its Applications}\
  ,\ \bibinfo {pages} {128714}} (\bibinfo {year} {2023})}\BibitemShut {NoStop}%
\bibitem [{\citenamefont {Liu}\ \emph {et~al.}(2014)\citenamefont {Liu},
  \citenamefont {Perra}, \citenamefont {Karsai},\ and\ \citenamefont
  {Vespignani}}]{Liu2014}%
  \BibitemOpen
  \bibfield  {author} {\bibinfo {author} {\bibfnamefont {S.}~\bibnamefont
  {Liu}}, \bibinfo {author} {\bibfnamefont {N.}~\bibnamefont {Perra}}, \bibinfo
  {author} {\bibfnamefont {M.}~\bibnamefont {Karsai}}, \ and\ \bibinfo {author}
  {\bibfnamefont {A.}~\bibnamefont {Vespignani}},\ }\href {\doibase
  10.1103/PhysRevLett.112.118702} {\bibfield  {journal} {\bibinfo  {journal}
  {Physical review letters}\ }\textbf {\bibinfo {volume} {112}},\ \bibinfo
  {pages} {118702} (\bibinfo {year} {2014})}\BibitemShut {NoStop}%
\bibitem [{\citenamefont {Perra}\ \emph {et~al.}(2012)\citenamefont {Perra},
  \citenamefont {Gon{\c{c}}alves}, \citenamefont {Pastor-Satorras},\ and\
  \citenamefont {Vespignani}}]{Perra2012}%
  \BibitemOpen
  \bibfield  {author} {\bibinfo {author} {\bibfnamefont {N.}~\bibnamefont
  {Perra}}, \bibinfo {author} {\bibfnamefont {B.}~\bibnamefont
  {Gon{\c{c}}alves}}, \bibinfo {author} {\bibfnamefont {R.}~\bibnamefont
  {Pastor-Satorras}}, \ and\ \bibinfo {author} {\bibfnamefont {A.}~\bibnamefont
  {Vespignani}},\ }\href {\doibase 10.1038/srep00469} {\bibfield  {journal}
  {\bibinfo  {journal} {Scientific reports}\ }\textbf {\bibinfo {volume} {2}},\
  \bibinfo {pages} {1} (\bibinfo {year} {2012})}\BibitemShut {NoStop}%
\bibitem [{\citenamefont {Moinet}\ \emph {et~al.}(2015)\citenamefont {Moinet},
  \citenamefont {Starnini},\ and\ \citenamefont
  {Pastor-Satorras}}]{Moinet2015}%
  \BibitemOpen
  \bibfield  {author} {\bibinfo {author} {\bibfnamefont {A.}~\bibnamefont
  {Moinet}}, \bibinfo {author} {\bibfnamefont {M.}~\bibnamefont {Starnini}}, \
  and\ \bibinfo {author} {\bibfnamefont {R.}~\bibnamefont {Pastor-Satorras}},\
  }\href {\doibase 10.1103/PhysRevLett.114.108701} {\bibfield  {journal}
  {\bibinfo  {journal} {Physical review letters}\ }\textbf {\bibinfo {volume}
  {114}},\ \bibinfo {pages} {108701} (\bibinfo {year} {2015})}\BibitemShut
  {NoStop}%
\bibitem [{\citenamefont {Baumann}\ \emph {et~al.}(2021)\citenamefont
  {Baumann}, \citenamefont {Lorenz-Spreen}, \citenamefont {Sokolov},\ and\
  \citenamefont {Starnini}}]{Baumann2021}%
  \BibitemOpen
  \bibfield  {author} {\bibinfo {author} {\bibfnamefont {F.}~\bibnamefont
  {Baumann}}, \bibinfo {author} {\bibfnamefont {P.}~\bibnamefont
  {Lorenz-Spreen}}, \bibinfo {author} {\bibfnamefont {I.~M.}\ \bibnamefont
  {Sokolov}}, \ and\ \bibinfo {author} {\bibfnamefont {M.}~\bibnamefont
  {Starnini}},\ }\href {\doibase 10.1103/PhysRevX.11.011012} {\bibfield
  {journal} {\bibinfo  {journal} {Physical Review X}\ }\textbf {\bibinfo
  {volume} {11}},\ \bibinfo {pages} {011012} (\bibinfo {year}
  {2021})}\BibitemShut {NoStop}%
\bibitem [{\citenamefont {Pierson}\ and\ \citenamefont
  {Schickler}(2020)}]{Pierson2020}%
  \BibitemOpen
  \bibfield  {author} {\bibinfo {author} {\bibfnamefont {P.}~\bibnamefont
  {Pierson}}\ and\ \bibinfo {author} {\bibfnamefont {E.}~\bibnamefont
  {Schickler}},\ }\href {\doibase 10.1146/annurev-polisci-050718-033629}
  {\bibfield  {journal} {\bibinfo  {journal} {Annu. Rev. Political Sci.}\
  }\textbf {\bibinfo {volume} {23}},\ \bibinfo {pages} {37} (\bibinfo {year}
  {2020})}\BibitemShut {NoStop}%
\bibitem [{\citenamefont {Leonard}\ \emph {et~al.}(2021)\citenamefont
  {Leonard}, \citenamefont {Lipsitz}, \citenamefont {Bizyaeva}, \citenamefont
  {Franci},\ and\ \citenamefont {Lelkes}}]{Leonard2021}%
  \BibitemOpen
  \bibfield  {author} {\bibinfo {author} {\bibfnamefont {N.~E.}\ \bibnamefont
  {Leonard}}, \bibinfo {author} {\bibfnamefont {K.}~\bibnamefont {Lipsitz}},
  \bibinfo {author} {\bibfnamefont {A.}~\bibnamefont {Bizyaeva}}, \bibinfo
  {author} {\bibfnamefont {A.}~\bibnamefont {Franci}}, \ and\ \bibinfo {author}
  {\bibfnamefont {Y.}~\bibnamefont {Lelkes}},\ }\href {\doibase
  10.1073/pnas.2102149118} {\bibfield  {journal} {\bibinfo  {journal}
  {Proceedings of the National Academy of Sciences}\ }\textbf {\bibinfo
  {volume} {118}},\ \bibinfo {pages} {e2102149118} (\bibinfo {year}
  {2021})}\BibitemShut {NoStop}%
\bibitem [{\citenamefont {Spencer-Rodgers}\ \emph {et~al.}(2010)\citenamefont
  {Spencer-Rodgers}, \citenamefont {Williams},\ and\ \citenamefont
  {Peng}}]{Spencer2010}%
  \BibitemOpen
  \bibfield  {author} {\bibinfo {author} {\bibfnamefont {J.}~\bibnamefont
  {Spencer-Rodgers}}, \bibinfo {author} {\bibfnamefont {M.~J.}\ \bibnamefont
  {Williams}}, \ and\ \bibinfo {author} {\bibfnamefont {K.}~\bibnamefont
  {Peng}},\ }\href {\doibase 10.1177/1088868310362982} {\bibfield  {journal}
  {\bibinfo  {journal} {Pers. Soc. Psychol. Rev.}\ }\textbf {\bibinfo {volume}
  {14}},\ \bibinfo {pages} {296} (\bibinfo {year} {2010})}\BibitemShut
  {NoStop}%
\bibitem [{\citenamefont {Noorazar}(2020)}]{Noorazar2020}%
  \BibitemOpen
  \bibfield  {author} {\bibinfo {author} {\bibfnamefont {H.}~\bibnamefont
  {Noorazar}},\ }\href {\doibase 10.1140/epjp/s13360-020-00541-2} {\bibfield
  {journal} {\bibinfo  {journal} {Eur. Phys. J. Plus}\ }\textbf {\bibinfo
  {volume} {135}},\ \bibinfo {pages} {1} (\bibinfo {year} {2020})}\BibitemShut
  {NoStop}%
\bibitem [{\citenamefont {Wang}\ \emph {et~al.}(2020)\citenamefont {Wang},
  \citenamefont {Sirianni}, \citenamefont {Tang}, \citenamefont {Zheng},\ and\
  \citenamefont {Fu}}]{Wang2020}%
  \BibitemOpen
  \bibfield  {author} {\bibinfo {author} {\bibfnamefont {X.}~\bibnamefont
  {Wang}}, \bibinfo {author} {\bibfnamefont {A.~D.}\ \bibnamefont {Sirianni}},
  \bibinfo {author} {\bibfnamefont {S.}~\bibnamefont {Tang}}, \bibinfo {author}
  {\bibfnamefont {Z.}~\bibnamefont {Zheng}}, \ and\ \bibinfo {author}
  {\bibfnamefont {F.}~\bibnamefont {Fu}},\ }\href {\doibase
  10.1103/PhysRevX.10.041042} {\bibfield  {journal} {\bibinfo  {journal} {Phys.
  Rev. X}\ }\textbf {\bibinfo {volume} {10}},\ \bibinfo {pages} {041042}
  (\bibinfo {year} {2020})}\BibitemShut {NoStop}%
\bibitem [{\citenamefont {Johnson}\ \emph {et~al.}(2020)\citenamefont
  {Johnson}, \citenamefont {Vel{\'a}squez}, \citenamefont {Restrepo},
  \citenamefont {Leahy}, \citenamefont {Gabriel}, \citenamefont {El~Oud},
  \citenamefont {Zheng}, \citenamefont {Manrique}, \citenamefont {Wuchty},\
  and\ \citenamefont {Lupu}}]{Johnson2020}%
  \BibitemOpen
  \bibfield  {author} {\bibinfo {author} {\bibfnamefont {N.~F.}\ \bibnamefont
  {Johnson}}, \bibinfo {author} {\bibfnamefont {N.}~\bibnamefont
  {Vel{\'a}squez}}, \bibinfo {author} {\bibfnamefont {N.~J.}\ \bibnamefont
  {Restrepo}}, \bibinfo {author} {\bibfnamefont {R.}~\bibnamefont {Leahy}},
  \bibinfo {author} {\bibfnamefont {N.}~\bibnamefont {Gabriel}}, \bibinfo
  {author} {\bibfnamefont {S.}~\bibnamefont {El~Oud}}, \bibinfo {author}
  {\bibfnamefont {M.}~\bibnamefont {Zheng}}, \bibinfo {author} {\bibfnamefont
  {P.}~\bibnamefont {Manrique}}, \bibinfo {author} {\bibfnamefont
  {S.}~\bibnamefont {Wuchty}}, \ and\ \bibinfo {author} {\bibfnamefont
  {Y.}~\bibnamefont {Lupu}},\ }\href {\doibase 10.1103/PhysRevLett.124.048301}
  {\bibfield  {journal} {\bibinfo  {journal} {Nature}\ }\textbf {\bibinfo
  {volume} {582}},\ \bibinfo {pages} {230} (\bibinfo {year}
  {2020})}\BibitemShut {NoStop}%
\end{thebibliography}%

\end{CJK}
\end{document}